\documentclass[sigconf]{acmart}

\AtBeginDocument{%
  \providecommand\BibTeX{{%
    \normalfont B\kern-0.5em{\scshape i\kern-0.25em b}\kern-0.8em\TeX}}}

\usepackage{algorithm, algorithmic}
\usepackage{amsmath}

\usepackage{xspace}
\usepackage{enumitem}
\usepackage{url}
\usepackage{xcolor,colortbl}
\usepackage{graphicx}
\usepackage{caption}
\usepackage{multirow}
\usepackage{booktabs}
\usepackage{subcaption}
\usepackage{amsmath,amsfonts}
\usepackage{graphics}
\usepackage{textcomp}
\usepackage{booktabs}
\usepackage{appendix}
\usepackage{enumitem}
\usepackage{xcolor}
\usepackage{colortbl}
\usepackage{float}
\usepackage{natbib}

\copyrightyear{2022}
\acmYear{2022}
\setcopyright{rightsretained}
\acmConference[UIST '22]{The 35th Annual ACM Symposium on User Interface Software and Technology}{October 29-November 2, 2022}{Bend, OR, USA}
\acmBooktitle{The 35th Annual ACM Symposium on User Interface Software and Technology (UIST '22), October 29-November 2, 2022, Bend, OR, USA}
\acmDOI{10.1145/3526113.3545678}
\acmISBN{978-1-4503-9320-1/22/10}
\begin{document}

\title{\sys: Screenshot-Based Bookmarks for Effective Digital Resource Curation across Applications}

\newcommand{\sys}{Scrapbook}
\newcommand{\capsys}{SCRAPBOOK}
\newcommand{\numberofparticipant}{13}
\newcommand{\numberofrsurveyparticipants}{125}

\newcommand\tabhead[1]{\small\textbf{#1}}
\newcommand{\trimtrimtrim}{\out{\vspace{-7pt}}}
\newcommand{\out}[1]{{#1}}

\newcommand{\sang}[1]{\out{{\small\textcolor{blue}{\bf [Sang: #1]}}}}
\newcommand{\donghan}[1]{\out{{\small\textcolor{purple}{\bf [Donghan: #1]}}}}
\newcommand{\TODO}[1]{\out{{\small\textcolor{red}{\bf [TODO: #1]}}}}
\newcommand{\TBD}[1]{\out{{\small\textcolor{yellow}{\bf [TBD: #1]}}}}
\newcommand{\revise}[1]{\out{{#1}}}

\newcommand{\marking}[1]{\out{{\small\textcolor{blue}{}}}}
\newcommand{\ADD}[1]{\out{\textcolor{black}{#1}}}

\makeatletter


\author{Donghan Hu}
\affiliation{%
  \institution{Virginia Tech}
  \city{Blacksburg}
  \state{Virginia}
  \country{USA}}
\email{hudh0827@vt.edu}

\author{Sang Won Lee}
\affiliation{%
 \institution{Virginia Tech}
 \city{Blacksburg}
 \state{Virginia}
 \country{USA}
 }
 \email{sangwonlee@vt.edu}


\begin{abstract}

Modern knowledge workers typically need to use multiple resources, such as documents, web pages, and applications, at the same time.
This complexity in their computing environments forces workers to restore various resources in the course of their work. 
However, conventional curation methods like bookmarks, recent document histories, and file systems place limitations on effective retrieval. 
Such features typically work only for resources of one type within one application, ignoring the interdependency between resources needed for a single task.
In addition, text-based handles do not provide rich cues for users to recognize their associated resources. Hence, the need to locate and reopen relevant resources can significantly hinder knowledge workers' productivity. 
To address these issues, we designed and developed \emph{\sys{}}, a novel application for digital resource curation across applications that uses screenshot-based bookmarks. 
Scrapbook extracts and stores all the metadata (URL, file location, and application name) of windows visible in a captured screenshot to facilitate restoring them later.
A week-long field study indicated that screenshot-based bookmarks helped participants curate digital resources.
Additionally, participants reported that multimodal---visual and textual---data helped them recall past computer activities and reconstruct working contexts efficiently.
  
\end{abstract}





\maketitle

\section{Introduction}
Modern knowledge workers need to carefully curate digital resources, such as files, applications, and web pages, since they often use multiple resources to do a single task~\cite{bannon1983evaluation, Hailpern:2011:YIR:1978942.1979165}.
For example, one may need to leave multiple windows open when writing an academic paper: a word-processing application, document viewers (for referencing other papers), web browsers (for searching for relevant information), and communication software (to collaborate with co-authors). 
Therefore, digital curation---discovering, organizing, searching, and retrieving such resources~\cite{jacobson2012information}---is an important daily task for knowledge workers. 
In particular, retrieving previously viewed and used digital resources, such as web pages, documents, local files, and software, is an inevitable task for knowledge workers~\cite{dragunov2005tasktracer, iqbal2007disruption}. 

However, current solutions for curating digital information have limitations. 
First, there is no single method that works for all types of resources. 
The resources needed for a single task may include files on a local computer, applications that run in a web browser, ordinary desktop applications (e.g., Terminal, Slack, Dictionary, or Outlook), and web pages containing information a worker intends to reference and retrieving them is different per each type. 
While functionality exists for users to curate each type of resource individually (e.g., desktop shortcuts, lists of recently opened documents, folder hierarchies, and browser bookmarks), their disparate nature forces workers to manage and restore resources of multiple types in a disjointed process. 
Second, identifying resources from largely text-based handles can be cognitively challenging. For example, a URL, a file name, or a web page title may not provide enough information for a user to recall the nature of the page's contents. 
It takes workers extra effort to locate and label each resource carefully (e.g., by renaming bookmarks and files or organizing folder hierarchies) to facilitate later recognition and navigation. 
Lastly, it is difficult to bundle heterogeneous resources from different applications into a single bookmark, so workers must recall and restore the relevant resources for each application separately. 
The burden of having to remember and reopen sets of resources is amplified by the length of task interruptions: a worker who wishes to pause work on a task for several days must either leave all the resources open until the next work session or recall and reopen them all several days later. 
Previous studies indicated that leaving windows open to eliminate retrieval tasks is a common solution, but doing this results in a cluttered computing environment and makes it more challenging to locate necessary resources among many open windows~\cite{ko2005eliciting, parnin2011resumption}. 

The online survey we conducted for our need-finding study (N=\numberofrsurveyparticipants{}), confirmed that knowledge workers are in need of effortless ways of curating digital resources across applications to improve their productivity.


\begin{figure*}[t]
  \centering
    \vspace{-10px}

  \includegraphics[width=\textwidth]{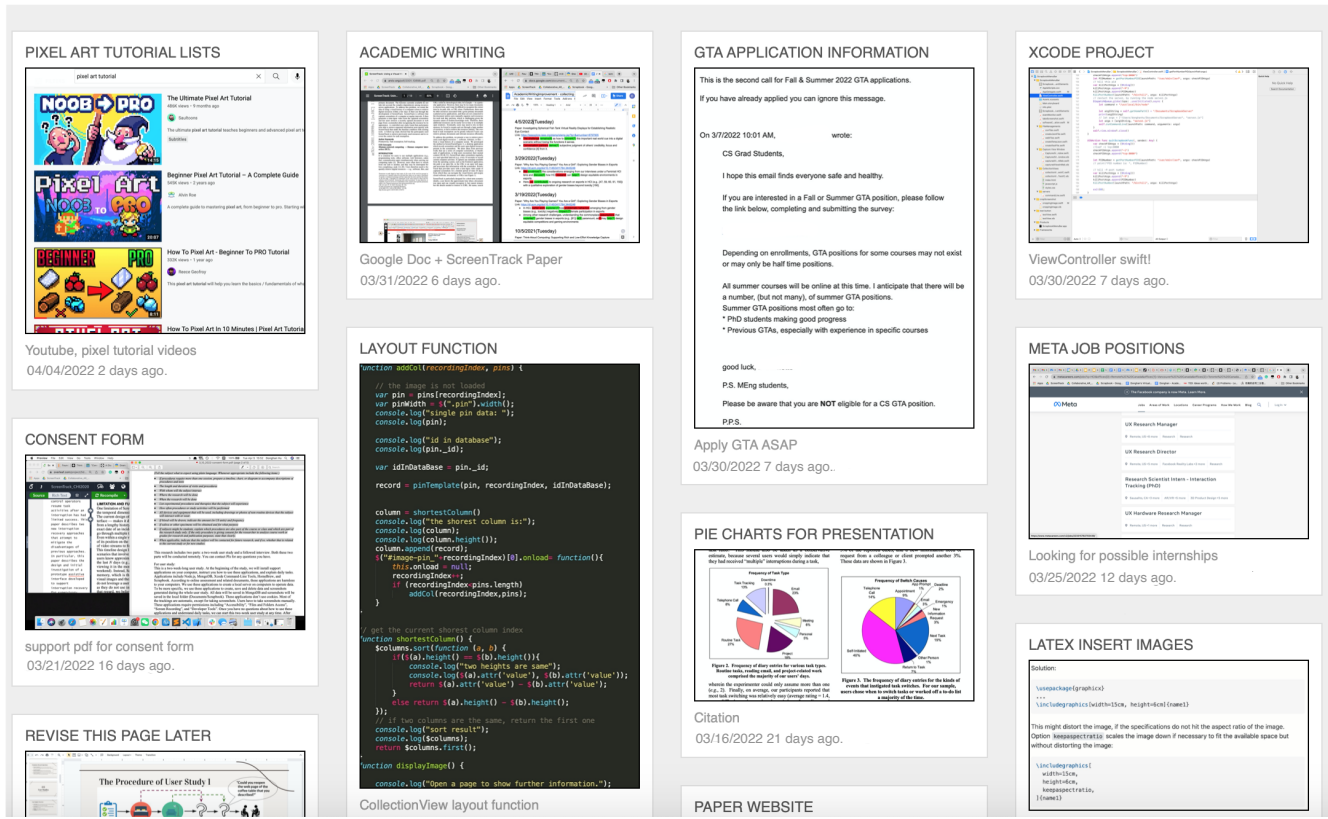} 
  \vspace{-20px}
  \caption{The Collection View in Scrapbook. The Collection View presents users with overviews of their previously captured computer activities. Here, screenshots are displayed in Masonry layout~\cite{Thornburg01} to roughly order them in a temporal order. For each capture, \sys{} presents a title, screenshot, description, and date so that users can have multimodal cues for recognizing the digital resources contained. Furthermore, users can enter keywords in the search field at the top to find specific captures. After locating a target screenshot, users can simply click on it to view further details and options for retrieval on the item's Detailed View page. See Figure~\ref{fig:twoViews}-(b) for The Detailed View.}
  \label{fig:collectionView}
  \Description{}
  \vspace{-10px}

\end{figure*}

To address this issue, we developed \sys{} (Figure~\ref{fig:collectionView}, a novel application that can capture any resources visible on a computer's screen when taking a screenshot. 
The goal of Scrapbook is to assist knowledge workers with the process of reconstructing their working context by allowing them to curate cross-application resources.
We use a screenshot as a handle to bookmark three types of resources: files, web pages, and applications. 
When a worker takes a screenshot, \sys{} will identify all the visible windows and extract meta-data from each one. 
We use a dynamic bit-masking algorithm to identify all the visible windows in a screenshot, thereby also capturing windows behind the frontmost application. 
Workers can bundle multiple applications into one screenshot bookmark by clicking a button, and users can later browse their collection of screenshots just as they might browse a photo album.
A screenshot provides rich visual cues to help workers better recognize the details of their past activities and other resources that they wanted to save for later, consistent with previous works~\cite{rule2017using, hu2020screentrack}. 
Lastly, \sys{} offers a search feature that allows users to search their screenshots with metadata, which may contain information like application names and URLs.  
  
To validate the effectiveness of Scrapbook in an ecologically valid setting, we conducted a one-week-long usability study with \numberofparticipant{} participants. 
The results indicated that Scrapbook helped users with the process of curating computing tasks and retrieving digital resources. The participants also found the Collection View helpful as a glanceable to-do list that helps remind them of their tasks.

The research contributions of this work are as follows:
\vspace{-5px}
\begin{itemize}[itemsep=0em,leftmargin=1em]
\item Design goals for digital resource curation methods, motivated by a formative study 
\item Exploration of the idea of screenshot-based bookmarks through the development of Scrapbook
\item Usability study results that demonstrate the effectiveness of screenshot-based bookmarks in retrieving digital resources
\end{itemize}
 
\noindent We believe that the screenshot-based method can serve as a novel method of curating digital resources with rich visual information and that this paper's results can motivate operating system--level curation methods that bundle cross-application resources for more effective curation.

\section{Related Works}
The ideas and motivation of \sys{} are drawn from prior works on (1) knowledge workers' challenges in multitasking and resource management, (2) metadata with visual and textual cues, and (3) recall and retrieval in human memory. 
In this section, we discuss the previous research in these domains to provide background for the design choices, features, and expectations for Scrapbook. 

\subsection{Users' Challenges in Resource Management}
Modern knowledge workers multitask routinely, with many applications, files, and web pages open at once. 
Gonzalez and Mark found that while modern knowledge workers deal with more than 10 distinct tasks per day, they spend an average of just 12 minutes on a particular activity before being interrupted~\cite{gonzalez2004constant}. 
A similar study stated that information workers usually spend less than one hour on a project before switching to another task~\cite{czerwinski2004diary}.
Given the inevitability of interruptions and frequent changes in working contexts, workers may encounter many resources of various types that they wish to curate, but curation activities---storing files, bookmarking web pages, opening files, and locating a web page---can degrade their working productivity.
For example, one study found that programmers usually spend more than one hour every day managing their working resources and projects to reduce time spent on recalling and retrieving them later~\cite{van1998interrupts}. 
Furthermore, spending time retrieving a past working context to restart an unfinished task can contribute to stress and negative emotions~\cite{czerwinski2004diary, mark2015focused}. 
Therefore, reducing the time spent curating resources would enhance computer users’ productivity.

\subsection{Tools for Organizing Digital Resources}
Curating digital resources on a computer involves various activities associated with organizing information and software, including seeking, discovering, organizing, using, and sharing. 
\ADD{Researchers found that information scraps in personal information management systems should be easy to capture and discover~\cite{bernstein2008information}.}
Digital curation activities can range from basic, low-level tasks like saving a file to complex, high-level tasks like creating and maintaining an organized hierarchy of nested folders.
Early HCI researchers focused on organizing multiple active windows based on tasks. 
For example, GroupBar allows users to manage windows in a batch grouped by task and expand and minimize batches of documents, which can be helpful for switching between tasks~\cite{smith2003groupbar}.
Scalable Fabric organizes thumbnail-like windows into groups near the edges of a computer's screen for workers to quickly switch between windows by clicking a thumbnail of the window~\cite{voida2008re}.
\ADD{Giornata similarly allows users to have a virtual desktop per activity~\cite{swearngin2021scraps}.}
TaskTracer applies machine learning algorithms to bundle various resources into a task~\cite{Dragunov:2005:TDE:1040830.1040855_tasktracer}.
\ADD{Some researchers distributed the effort needed to curate resources by inserting microtasks when users are interrupted~\cite{teevan2014selfsourcing}. 
Inserting such microtasks was helpful for task resumption as well~\cite{williams2019mercury}. 
Capturing context and transferring resources across devices can be challenging, as workers may use a variety of devices for a task~\cite{swearngin2021scraps}. }\marking{1} 
\ADD{Furthermore, to manage and handle scattered digital information, researchers have proposed various technical methods to facilitate task-based computing activities and improve the efficiency of microproductivity~\cite{ bernstein2007management, shen2008automatically, oleksik2009lightweight}.}\marking{1}
\ADD{These approaches provide effective means of bundling or curating task resources in the context of window management for multitasking and cross-device work support, assisting users in quickly switching between tasks and curating resources on the go.} 

In contrast, \sys{} provides tools for knowledge workers to curate digital resources for various purposes in a longer term with visual cues: bookmarking an article, initializing a computing environment for task resumption, saving multiple resources related to a given resource, or keeping track of unfinished tasks for mental reconstruction. 

With the proliferation of web-based resources and applications, researchers have developed many tools for managing web resources.  
One native, widely used tool in modern web browsers is the bookmark function. 
However, researchers have found that as a user's number of bookmarks increases, browsing, locating, and retrieving web pages becomes more challenging, ultimately requiring organizational efforts comparable to those applied to file systems: users must establish a hierarchical structure of bookmark folders, name bookmarks carefully, and categorize them immediately upon creation~\cite{10.1145/274644.274651}. 
\ADD{Cockburn et al. found that web users maintain large collections of bookmarks, nearly 200 on average, thereby yielding a lengthy list of bookmarks with folders~\cite{cockburn2001web}. 
The authors found empirical evidence demonstrating that users do not carefully curate bookmarks: imbalances between deletion and addition lead to the appearance of duplicate and invalid bookmarks.}
The limitations of native implementations of bookmarking may account for the popularity of commercial apps like Pocket~\footnote{\url{https://getpocket.com/}} and Pinterest~\footnote{\url{https://www.pinterest.com}}, which enable social bookmarking and provide multimodal cues (images and text) to facilitate web content curation. 
Similarly, collaborative web browsing and bookmarking was shown to be an effective curation method for a specific task~\cite{10.1145/1294211.1294215}.
Recently, researchers have developed ways to organize multiple tabs based on search keywords~\cite{10.1145/1357054.1357242}, themes~\cite{10.1145/3173574.3173825}, or tasks~\cite{10.1145/3472749.3474777}. 
While these tools facilitate effective resource curation, they do not allow workers to curate resources across applications.

\subsection{Using Visual Cues and Screenshots to Reconstruct a Working Context}
\label{sec:visual}
Researchers have found that human beings can remember visual information at scale~\cite{brady2008visual}.
Standing et al. found that humans were able to determine whether they had seen an image two days ago with 90\% accuracy among a collection of 2560 pictures~\cite{standing1970perception}. 
Sellen et al. ran a case study and found that humans could correctly identify images of their own among those of others with over 80\% accuracy in life-logging context~\cite{sellen2007life}.
In addition, visual information can be processed in parallel, meaning that a human can recognize an image at a glance; conversely, humans cannot recognize a block of text visually and must parse it for meaning~\cite{WARE20211}. 
These results indicate that humans are able to discriminate between similar images, recall detailed contexts, and remember related past activities from images.

The human ability to recall details from visuals has often been used in the context of computing as well. 
In one study, Czerwinski and Horvitz stated that video footages of a worker using a computer serve as effective cues for people to recall past events even after one month~\cite{czerwinski2002investigation}. 
The recall capability by seeing computer recording was developed into a research method---usability testing with retrospective recall as an alternative to the think-aloud technique~\cite{russell2009retrospective, russell2014looking}.
Application windows serve as visual cues for knowledge workers to understand which (sub)tasks need to be finished when workers resume a task~\cite{iqbal2007disruption, ko2005eliciting}.
In programming environments, researchers implemented methods to display snippets of edited code blocks to help programmers resume past coding and debugging tasks~\cite{safer2007comparing, parnin2010evaluating}.
Modern smartphone users take advantage of the ubiquity of smartphone cameras with the practice of taking photos to serve as external memory devices; for instance, a user might take a photo of a pillar to recall their car's location in an airport parking lot or take a screenshot of mobile phone to save a QR code of a flight ticket~\cite{Finley2018}.
In this work, the metaphor of photo album was used for the Collective View in \sys{} design.

Many HCI researchers have leveraged the human ability to comprehend visual information in creating productivity tools for mental reconstruction and task resumption. 
For example, \revise{Taskpos{\'e} arranges thumbnails of windows based on their relevance to an ongoing task, helping users to manage multiple windows based on usage patterns and transitions between them~\cite{Bernstein:2008:TEF:1449715.1449753}.}
A precursor to this work is WindowScape, which allows users to capture a group of windows on a screenshot to switch between digital work spaces~\cite{10.1145/2147783.2147791}. 
However, these systems only allow transitions between already open windows; consequently, they do not address the challenge of long-term resource curation and retrieval that workers face. 
Another application of screenshots explored in numerous studies lies in creating a visual history to help users recall the mental context of their computing activities in the past.
Rule et al. investigated the benefits of using a visual history for mental reconstruction and arrived at design implications by exploring various design options (animation vs. still image, thumbnail size) and their effects~\cite{rule2015restoring, rule2017using}.
ScreenTrack generates a time-lapse video by capturing screenshots periodically; workers can reopen the frontmost application from any frame of the video and use the video to reconstruct their mental context~\cite{hu2020screentrack}. 
While these systems and the present work share the goal of facilitating mental reconstruction, they differ from \sys{} in as much as they visualize screenshots in the temporal dimension rather than using them for explicit curation

\section{Formative Study}
We conducted a formative study to understand potential challenges and common strategies in curating digital resources for modern knowledge workers. 
Specifically, we wanted to understand computer users' attitudes toward traditional curation methods. 

\subsection{Participants and Survey}
We designed an online questionnaire to investigate what computer users think of current curation methods and personal task resumption methods.
We created an online survey with questions in the following categories: (1) features that users use for web pages and file curation, (2) challenges for task resumption, and (3) demographic questions.
The survey included five-point scale items prompting users to report their feature usage frequency, as well as open-ended questions. 
Respondents were recruited through a variety of channels: the author's social media accounts, Amazon Mechanical Turk, and campus mailing lists.
In each case, eligibility criteria posted with the survey indicated that respondents should be modern computer users who use computers in their work.
In total, we received 125 responses from these platforms after excluding those responses that did not pass attention check questions. 
Each participant received a \$5 e-gift card for their participation.
Respondents had an average age of 32.6 years; 63 were male, 60 were female, and 2 chose not to provide their gender. 
Out of the 125 respondents, 115 had at least some form of college degree, and 118 rated their digital literacy as average or above. 
Participants reported spending an average of 8.6 hours per day using computers for daily tasks. 

\begin{figure}[t]
\centering
  \includegraphics[width=\columnwidth]{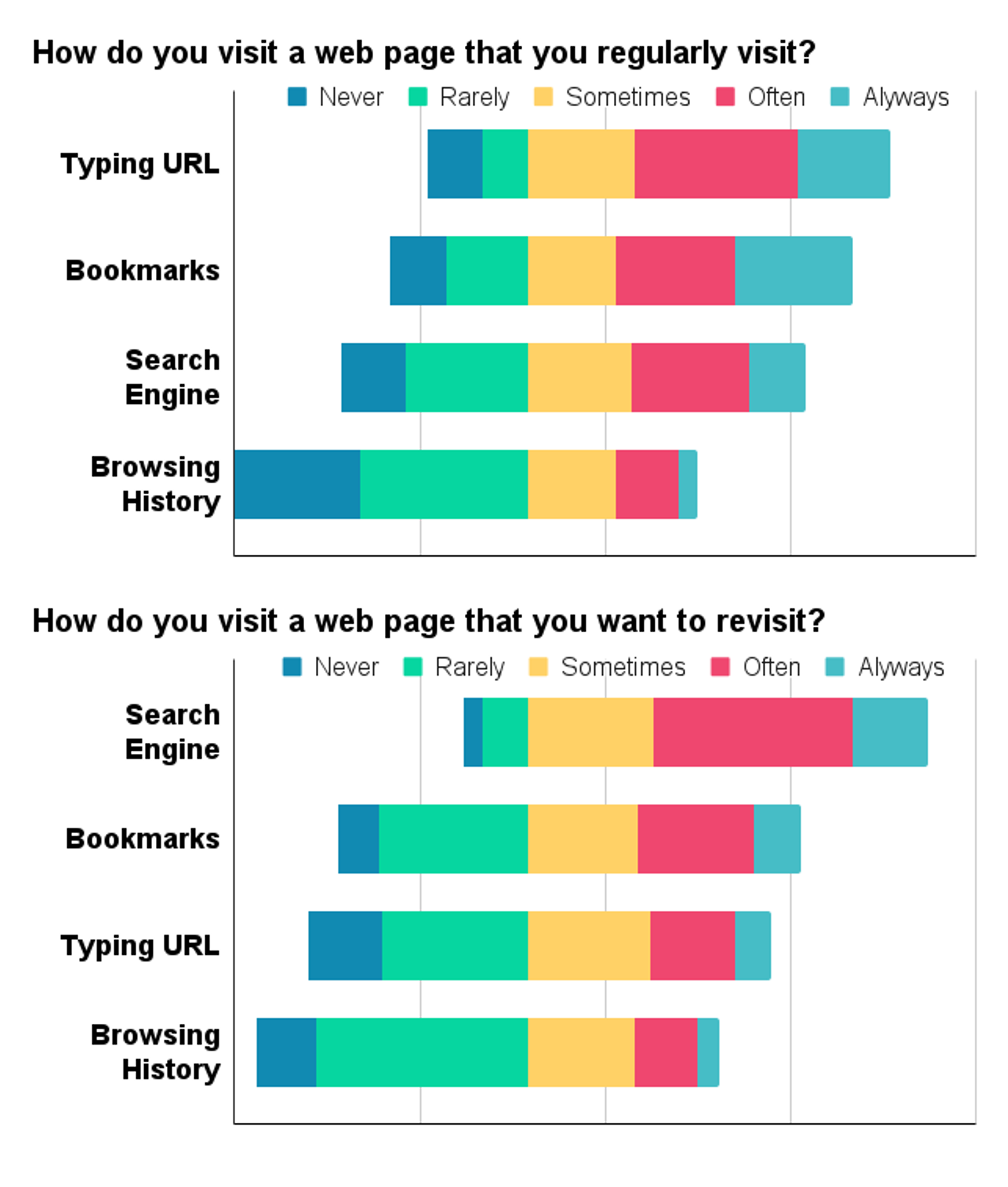} 
  \caption{Participants' ratings of how frequently they use different methods to revisit a website. The upper bar graph illustrates the proportion of reported usage frequencies for the listed methods of opening regularly visited websites. The lower graph illustrates the proportion of reported usage frequencies for the listed methods of reopening needed websites that respondents did not visit regularly.}
  ~\label{fig:bookmarks}
\end{figure}
\subsection{Results}
\subsubsection{Bookmarks are infrequently utilized.}

We calculated descriptive statistics for the five-point scale items, and the first author conducted a thematic analysis of the open-ended responses to generate codes. The codes and themes were reviewed by both authors and iteratively coded the responses until common themes emerged from the data.

We asked participants how often (``never,'' ``rarely,'' ``sometimes,'' ``often,'' or ``always'') they used built-in web browser functionality to visit websites in two categories: regularly visited sites (such as YouTube or Amazon) and sites they do not visit regularly but wish to revisit. 
The results indicate that the resource management features web browsers provide---namely, bookmarks and a browsing history---are not the most frequently used methods for revisiting websites. 
For regularly visited websites, users tend to simply type their URLs in the address bar; for irregularly visited sites, users repeat keyword searches to find the sites again (see Figure~\ref{fig:bookmarks}.).
Bookmarks are the second most used function in both cases.  
While Bookmarking would be the function dedicated to resource curation, it is not the most frequently used method of revisiting websites, being the less-used method than relying on their memory (URLs, keywords) or searching it from scratch (search engine).  
This result is consistent with a previous work which found that history and bookmarking mechanisms provided by browsers can be ineffective~\cite{10.1145/274644.274651, weinreich2006off}.

We further asked about the inconvenience of using bookmarks in web browsers in an open-ended question. 
The most common answers (38/125) mentioned that curating bookmarks is time-consuming and bookmarks are ``cumbersome to keep organized (P52).'' 
Participants complained about having to give bookmarks recognizable names and place them in the right folders. 
Similarly, some participants (21/125) complained about the difficulty of finding a desired bookmark, especially when bookmarks lack descriptive labels. 
Other responses were associated with the scale of bookmark collections, pointing out the challenge of locating a desired bookmark within a large field of irrelevant ones.
They pointed out that they simply had too many bookmarks (20/125), and sometimes, they even forgot about what pages they had bookmarked (18/125).
These responses clearly show that bookmarks are not scalable from a certain point. 
Evidently, having to organize bookmarks to later recognize them easily is the biggest challenge in using bookmarks. 






\subsubsection{Users get lost in folder hierarchies.}

Survey respondents most commonly use their operating system's native file manager (e.g., Finder, Windows Explorer) to curate files; 70 out of 125 chose Often or Always. 
Arranging files on the desktop was also identified as a common approach (Often/Always: 58 out of 125) to saving frequently accessed documents and files.
Other less commonly used features were recent documents and search features (e.g., Spotlight, Everything). 
Search features were used especially when users could not locate files by other means. 

Mirroring the situation with bookmarks, the biggest concern respondents had regarding retrieving files was not being able to locate files or paths to files (48/125).  
P16's response exemplifies the challenges of remembering file locations:
\begin{quote}[P16]
"When I'm trying to find a file and don't remember where I saved it, I'll type in what the title is and sometimes I can't remember what the document is titled." 
\end{quote}
\noindent In addition, recognizing poorly named files (19/125) and navigating a folder hierarchy (18/125) were named as challenges. 
P20's response demonstrates the cost of organizing files immediately upon saving them, which can impact a later retrieval task: 
\begin{quote}[P20]
"I name files something that makes sense at the time while I'm in a rush, but then have trouble recalling where I put it or what that name means." 
\end{quote}
\noindent Lastly, while we did not specifically check the portion of users using tools other than file systems, we realize that users have a variety of choices, which increases the number of platforms involved in curation: folder hierarchies, desktops, cloud file systems (e.g., Google Drive and Dropbox), bibliography software, browser bookmarks, and code repositories, to name a few. 


\subsubsection{Multiple Challenges: Recall, Locating, and Retrieval}
\label{sec:multiple:challenges}
We asked the participants about the kinds of challenges they face in restoring working contexts in an open-ended question. 
Many participants (58/125) reported that the need to track where they left off when they resume a task is their biggest concern---to reconstruct their mental context. 
Thus, knowing the context of what needs to be retrieved is the biggest challenge. 
The second most common theme (38/125) was retrieving resources as a challenge in resuming a task. 
In addition to these challenges, many participants pointed out the relationship between resource retrieval and the burden of organizing resources: resources collected haphazardly are challenging to locate later.
\begin{quote}[P70]
"The biggest challenge is finding it again. Having things organized takes forethought and time."
\end{quote}
 \begin{quote}[P32]
"Having multiple versions of a file saved and remembering which one I was working on because I named them inefficiently."
\end{quote}
\noindent To address these challenges, many participants (56/125) responded that they often leave windows open. 
In general, leaving too many applications open can impact a computer's performance and create visual clutter, ultimately making it more challenging for users to locate desired resources.

\section{\sys{}: Design and Implementation}
Drawing ideas from screenshot-based productivity tools ~\cite{10.1145/2147783.2147791, rule2017using, hu2020screentrack}, we designed and implemented \sys{}. 
In this section, we introduce the design components that we set to address the challenges identified in the formative study. 
We also introduce an implementation for effective curation and task resumption.
In particular, we consider various actions inherent to curating digital resources---recording, browsing, searching, and retrieving---and we explain the design of \sys{} as it relates to these actions. 


\subsection{Screenshot-Based Bookmarking for Cross-Application Resources}
In \sys{}, we use a screenshot as a handle to store digital resources in a bundle. 
The items visible on a computer screen have already been used as a visual cue to help workers reconstruct their mental context~\cite{iqbal2007disruption, ko2005eliciting, safer2007comparing, parnin2010evaluating}.
One benefit of using images of a computer screen is that all software with a graphical user interface (GUI) inherently affects what is shown on the screen, enabling a curation method for cross-application resources.  
While existing works often capture an active window on a screenshot~\cite{bernstein2008taskpose, hu2020screentrack, ScalableFabric}, this approach is limited to one window or the frontmost application, thereby excluding resources underneath the frontmost window that may be relevant to the task.
In practice, even when one is under the other in terms of its order, two windows can be laid out side by side so that both windows are fully visible.
In \sys{}, a user can bundle multiple resources that are commonly used together. 
Lastly, we anticipate that users will quickly recognize the contents of a screenshot at a glance and recall the context in which they took a screenshot based on the human capability to process visual information introduced in~\ref{sec:visual}.
\ADD{The benefit of visual cues will alleviate the challenge users face in reconstructing their mental context identified in the formative study, and the simple interaction of capturing a screenshot will reduce the burden of curating resources.}\marking{2}

\begin{figure}[t]
  \centering
  \includegraphics[width=0.80\columnwidth]{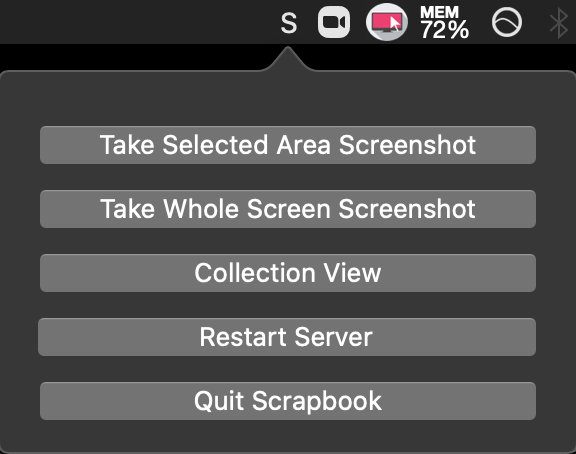}
  \caption{The Scrapbook popup menu. This menu pops up when users click the "S" icon in the menu bar. Users can take screenshots with the first two buttons to capture their current computer activities. Clicking the Collection View opens the UI shown in Figure~\ref{fig:collectionView}.}
  \label{fig:MenuBar}
  \Description{} 

\end{figure}

\sys{} provides two screenshot capture options: capturing a screenshot by selecting an area (the first button in Figure~\ref{fig:MenuBar}) and taking a full-screen screenshot (the second button in Figure~\ref{fig:MenuBar}). 
The former is intended for use in targeting a specific application or a specific region of a screen (e.g., a figure in an academic paper).
These options resemble the capture options available in an operating system's native screenshot functionality. 
\ADD{We anticipate that the simplicity and universality of curating any visible resource with a single click will help to reduce knowledge workers' burden of expressly choosing a curation method per resource type.}\marking{3}

\subsection{Extracting Metadata per Resource}

\sys{} extracts all available metadata from each window visible in a screenshot. 
While a screenshot may help a user reconstruct their mental context, a screenshot alone cannot open a file, an application, or a web page depicted therein. 
Therefore, metadata, such as file names, URLs, file paths, and application paths, is essential in reopening applications. 
We use an operating-system-level accessibility API to extract metadata.
\ADD{Combined with a screenshot, this metadata practically eliminates the challenge of locating resources immediately after recalling them from an image, which was one of three challenges we found in the formative study. }

In addition to the metadata mentioned above, \sys{} collects other information that may assist users when using the search feature, such as application names, web page titles, window titles (which may not simply be the names of open files), or any user-provided data they may choose to add.  
For instance, a user may be able to resume reading a paper by searching for its title and finding a PDF viewer window whose title contains the paper's title. 
Therefore, \sys{} will provide \textit{multimodal}---visual, temporal, and textual---cues that users can rely on to locate a screenshot, which will be discussed in ~\ref{fig:collectionview}. 


\subsection{Identification of Visible Captured Applications}

\begin{figure}[t]
  \centering
  \includegraphics[width=\columnwidth]{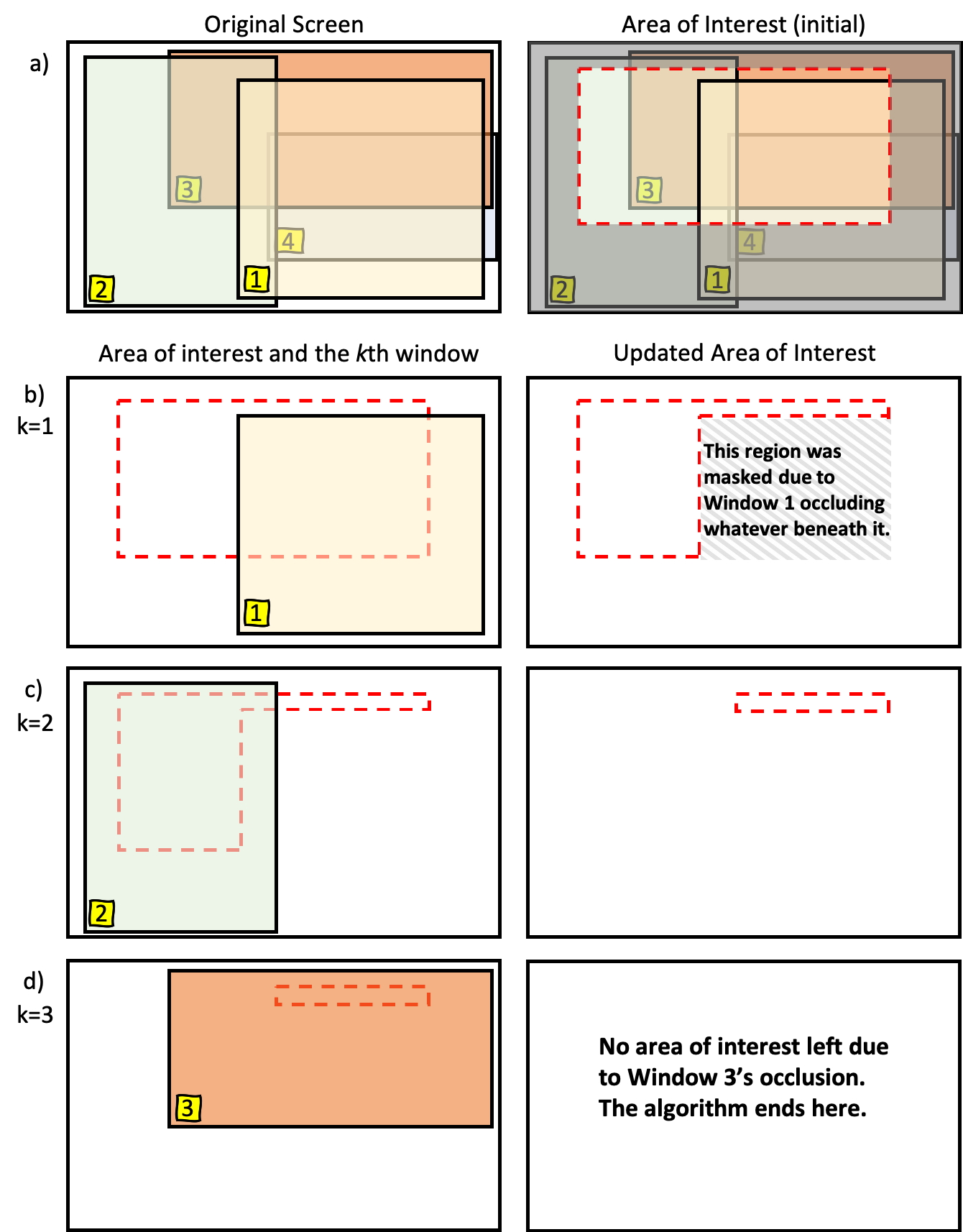}
  \caption{Step-by-step illustration of the bit-masking algorithm. (a) Suppose there are four windows 1--4 in that order (left) and a user takes a selected area screenshot (right; selected area of interest highlighted in red). (b) The algorithm sees the intersection between Window 1 and the selection to determine the visibility of the window within the screenshot (left). If visible, it updates the area of interest by excluding the intersection, as any following windows in the area will be occluded by Window 1. (c) Given the updated area of interest, examine the next window available until there is no window to examine or the area of interest is too small.}
  \label{fig:algorithm}
  \Description{Algorithm Demonstration}
\end{figure}

\sys{} identifies visible windows from a screenshot using a bit-masking algorithm. 
The appearance of a screenshot depends in part on how a user arranges their windows. 
For example, in a simple case, a user may use only one window at a time in full-screen mode. 
Conversely, some users may use ultra-wide monitors or arrays of several monitors to view multiple large windows at once.
In a more complicated case, a user may want to include a partially visible window in a screenshot, such a window header of a relevant resource that is partially visible behind the main resource window. 
A complex example is shown in Figure~\ref{fig:algorithm}-a; note that all four windows in the example are, at least partially, visible. 
While \sys{} can extract metadata regarding every window's location --- its width, height, the coordinates of its upper-left corner, and its ``depth'' among other windows, it does not explicitly determine what portion of any window is visible. 
The visible region of Window 3 in Figure~\ref{fig:algorithm}-a, for example, is an irregular, concave octagon, which makes it difficult to evaluate whether the window is visible or not by simply knowing its size and location; it depends on where the windows above it (1 and 2, in this case) are located. 
The simplest possible solution is to capture only the frontmost application, but this would prevent our tool from capturing multiple resources. 
Capturing all open applications is another option, but this may force a user to remove/deselect irrelevant windows when users retrieve resources from a screenshot. 
Therefore, we identify the visibility of each window by running a bit-masking algorithm (see Figure~\ref{fig:algorithm} and Algorithm~\ref{Algorithmpseudocode}). 

\begin{algorithm}[t]
\caption{\textcolor{black}{Identifying visible windows on a screenshot.} Matrices are denoted by italicized capital letters. Non-italicized capital letters use set notation. The dimensions of all matrices are the dimensions of the computer screen after downsampling.} 
	\label{Algorithmpseudocode}
 	\begin{algorithmic}[1]
	    \STATE let {$A$} = [$a_{(i,j)}$], where \COMMENT{\textcolor{purple}{$A$: Area of Interest Matrix}}
	    \[a_{(i,j)}=\left\{\begin{array}{cl}
1,& \mbox{if (i,j) in screenshot area}\\
0,& \mbox{elsewhere}\end{array}\right.\]
        \STATE let V = \{ \}  \COMMENT{\textcolor{purple}{set of visible window index, empty set initially}}
		\FORALL{Window$_k$=1,2, $\ldots$, n  \COMMENT{\textcolor{purple}{n: the \# of windows on screen}}}
		
        \STATE let {$W_k$} = [$w_{k,(i,j)}$], where \COMMENT{\textcolor{purple}{$W_k$: the \textit{k}th Window Matrix}}
 	    \[w_{k,(i,j)}=\left\{\begin{array}{cl}
 1,& \mbox{if (i,j) in the $k$th Window area}\\
 0,& \mbox{elsewhere}\end{array}\right.\]
            \STATE let {$O_k$} = [$o_{k,(i,j)}$], where \COMMENT{\textcolor{purple}{(O$_k$ = W$_k$ $\cap$ A)}}
            	    \[o_{k,(i,j)}=w_{k,(i,j)} \&  a_{(i,j)}\]
    	    \IF{$elementSum(O_k) > threshold$} 
    	        \STATE V.add(k) \COMMENT{\textcolor{purple}{the $k$th Window is visible.}}
	        \ENDIF
	        \STATE update {$A$} = [$a_{(i,j)}$], where \COMMENT{\textcolor{purple}{(A = A - W$_k$)}}
            	    \[a_{(i,j)} \gets a_{(i,j)} \&  \neg w_{(i,j)}\]
    	    \IF{$elementSum(A) \leq threshold$}
                \STATE break; \COMMENT{\textcolor{purple}{Area of interest smaller than threshold}}
	        \ENDIF
	        
		\ENDFOR\\
		\RETURN V \COMMENT{\textcolor{purple}{Return the set of visible window index}}
 	\end{algorithmic} 
\end{algorithm}

The core idea of the bit-masking algorithm is to dynamically maintain a matrix, denoted by A in Algorithm~\ref{Algorithmpseudocode}, that represents an area of interest considering the occlusion between windows. 
For example, initially, each element $a_{(i,j)}$ of matrix A is 1 if the element is within the screenshot area (Figure~\ref{fig:algorithm}-a-Right; line 1 in Algorithm~\ref{Algorithmpseudocode})---if a user takes a full-screen screenshot, A will be a matrix of ones.
Then, the algorithm iterates over each open window from the frontmost to rearmost to see if there is any intersection between the area of interest and a window area (lines 4 and 5). 
If there is an intersection and the intersection is greater than a threshold value (line 6), the target window is determined to be (sufficiently) visible in the screenshot (Figure~\ref{fig:algorithm}-b-Left, line 7). 
Next, the area of interest is updated before moving on to the next window to exclude the region of the most recent visible window; whatever is underneath the window will be invisible. (Figure~\ref{fig:algorithm}-b-Right, line 9)
The algorithm repeats the intersection check (Figure~\ref{fig:algorithm}-b, c, d, Left) and updates the area of interest (Figure~\ref{fig:algorithm}-b, c, d, Right) until there is no window left on the screen or the area of interest is smaller than a threshold value (Figure~\ref{fig:algorithm}-d-Right, line 10-12). 
We reduced the computational load of the algorithm by downsampling the screen and all window sizes/locations. Also, by setting the threshold value, we can filter out small windows that are unlikely to contain meaningful information. 
If the downsampling ratio is 10 pixels in length and the threshold is 5, the algorithm will ignore any window whose visible area is less than 500 (e.g., a 50~×~10 area).
In the example of Figure~\ref{fig:algorithm}, the algorithm might have ended after Window 2 if the area of interest (e.g., Figure~\ref{fig:algorithm}-c-Right) were smaller than the threshold. 
We used 0 for the threshold value for the sake of simplicity; users would be able to understand that \sys{} will capture any resource that is visible on a screen, no matter how small the visible part is.  
The visibility identified per window will be used to recommend which application to open, discussed in~\ref{sec:detailedView}.
\begin{figure}
    \begin{subfigure}[b]{\linewidth}
        \centering
              \vspace{-10px}

        \includegraphics[width=\linewidth]{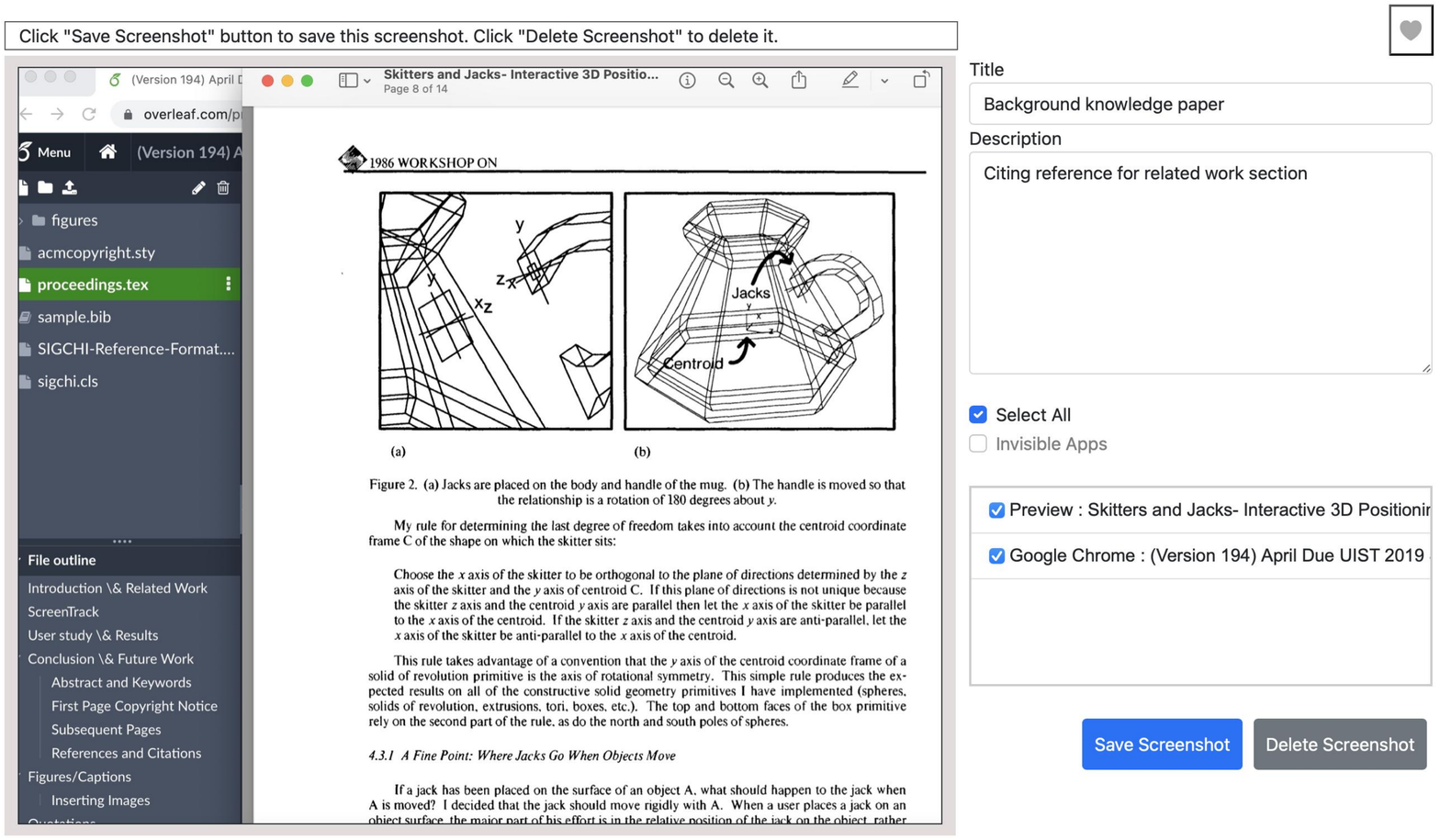}
        \caption{Captured View}
    \end{subfigure}
    \hfill
    \begin{subfigure}[b]{\linewidth}
        \centering
        \includegraphics[width=\linewidth]{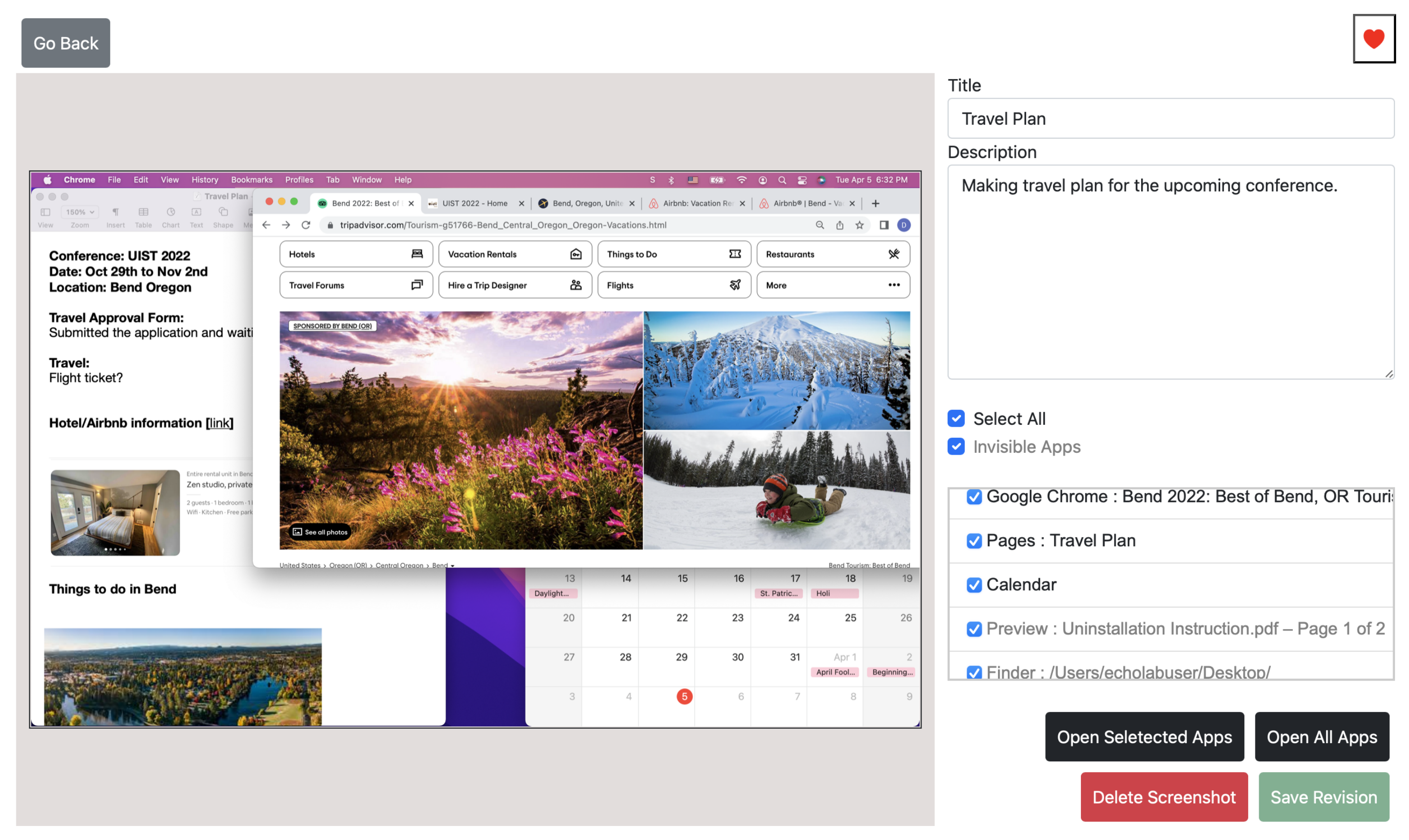}
        \caption{Detailed View}
    \end{subfigure}
\caption{(a) The Captured View appears after a user takes a screenshot to curate their resources; in this case, a PDF of an academic paper and an Overleaf project. (b) In Detailed View, users can view more details about a task. The example shows the task of planning a trip across multiple applications.}\label{fig:detailedView}
\label{fig:twoViews}
\end{figure}

\subsection{Browsing a Screenshot Collection}
\label{fig:collectionview}
The design used to present curated screenshots may influence the effectiveness of \sys{}. 
For example, there is a trade-off between the screenshot size and the number of images that can be displayed within a full-screen window simultaneously. 
Conversely, larger thumbnails can promote more accurate mental context recall~\cite{rule2017using}. 
We chose to provide two different views: the Collection View, which shows a collection of screenshots; and the Detail View, which shows a single screenshot. 
\ADD{The Detail View assists users in the challenge of recognizing content from its bookmark handle with rich, visual cues available in a screenshot; conversely, the Collection View reduces the cost to users of having to organize bookmarks by linearly presenting them like a photo album.}\marking{2}
We will share the Collection View design in this section and Detail View in the following section. 

Users can access their curated screenshots in the Collection View window~\ref{fig:collectionView}.
The Collection View is the core component that helps users glance at their collections and locate screenshots. 
Whenever a user wants to reopen an application from a screenshot or immediately after a user saves a screenshot, the user is routed to the Collection View, providing an opportunity to glance at the collection and reminding them of recently curated screenshots.   
The Collection View offers three ways to search for screenshots, in addition to screenshots' visual cues. 
The most basic method is skimming screenshots in the temporal dimension by scrolling the screenshot feed. 
The Collection View uses a ``Masonry'' layout, patterned on online systems for curation like Pinterest, Google Keep, and photo albums~\cite{Thornburg01}.
Overall, the screenshots are presented in reverse chronological order, with the most recent one in the upper-left corner. 
Users can understand the ordering of screenshots through timestamps presented as the time distance between the capture time and the current time (e.g., 3 days ago). 
The second way is to search by keywords. 
A search bar is located at the top of the interface; users can search for screenshots whose metadata matches their input in any of the following fields: filename, application name, date, URL, window title, and user-supplied data (such as a description or title).
Searching by keyword is helpful in drastically reduce the number of screenshots users must skim, if not to locate the target screenshot immediately. 
Lastly, users can like---by pressing heart button---frequently used screenshots to make them appear before others.
This feature is similar to marking important emails with a star, flag, or pin in an email client~\cite{10.1145/2556288.2557013}. 
If a user wants to capture the work context of an interrupted task that they want to resume in a short term, starring the screenshot will be useful. 

\subsection{Capturing Detail per Screenshot}
\label{sec:detailedView}

Lastly, retrieving resources from a screenshot is a key feature of \sys{}. 
Immediately after a user takes a screenshot, the user is given the opportunity to add their own input and select or deselect resources captured from the screenshot in the Capture View (Figure~\ref{fig:detailedView}-(a)).
\ADD{This function helps users contextualize their curation beyond images and captured metadata, similarly to a previous work~\cite{swearngin2021scraps}.
}\marking{2}
\sys{} selects all of the applications initially evaluated as visible by the bit-masking algorithm. 
Users have an option to deselect visible applications if they are irrelevant to the task. 
Similarly, users can also include invisible apps in the screenshot by expanding the invisible apps and choosing an application to add. 
This functionality will be useful in situations where a user switches between two full-screen windows, one of which will always be fully occluded by the other, despite using two resources simultaneously (e.g., Google Scholar and Microsoft Word). 
In addition, including invisible resources can facilitate mental reconstruction. 
For example, a user may wish to curate an article they encountered (\textit{a curated resource}) while writing a paper about social media (\textit{context}), which might have been invisible on a screenshot. 
By including this information, users can understand the context in which they decided to curate a given resource. 
Lastly, a user can optionally supply their own information in the Title and Description fields, which may be useful for later searches. The title defaults to the time when the screenshot was taken. 
Generally, clicking Save Screenshot will be sufficient if a user simply wants to save the resources visible in the screenshot. 
Once saved, \sys{} will display the Collection View to show the most recent screenshot added.

Users can reopen captured resources from the Detailed View (Figure~\ref{fig:detailedView}-(b)) for a screenshot, accessed by clicking any thumbnail in the Collection view. 
\sys{} will recommend which applications be opened 
Users can choose to open more or less than initially selected by the algorithm, including invisible resources. 
In both the Capture View and the Detailed View, images are displayed at maximum size with an option to zoom in to see the original size. 
In addition, each resource's metadata (e.g., the application name, window title, and path) will be presented. 
 
\subsection{Implementation}
We designed and implemented \sys{} for Apple's macOS.
The program uses AppleScript to extract metadata from captured applications \ADD{and relaunch applications}.
The process of running AppleScript code and capturing screenshots is accomplished using macOS's native accessibility API.
\ADD{AppleScript supports identifying and opening a specific view in an application with customized per-application scripts, such as a website in browser, a document in Microsoft Word, a project in programming software, or a particular event in a calendar.
For instance, if users capture a screenshot for a document displayed in Microsoft Word, Scrapbook will run predefined AppleScript code based on the captured application and extract metadata, including the file's name and path.
Later, when a user wants to reopen the document, Scrapbook will run additional AppleScript code to launch Microsoft Word and open the target file from the stored local path.}\marking{4}

\ADD{Scrapbook can curate any application and identify corresponding digital resources, provided that the application supports scripting.
In the current design, we mainly consider productivity tools and frequently used applications to cover typical computing tasks, including browsers, the Microsoft Office suite, Apple's iWork suite, programming software, PDF readers, file managers, and text editors. 
Furthermore, we designed the software architecture so as to facilitate remotely updating the AppleScript code by maintaining an external CSV file containing customized, application-specific scripts.
}\marking{4}
All metadata and screenshots are stored in a local MongoDB database.
We used Node.js, an open-source, back-end JavaScript runtime environment, to set up a local server for presenting user interfaces, screenshots, and metadata, and to provide a retrieval method in the browser interface.
We developed \sys{} as a window-less application that runs in a menu bar, allowing users to take a screenshot without having to locate the window (Figure~\ref{fig:MenuBar}).


\section{Evaluation: A Week-Long Usability Study}
We conducted a one-week field study to evaluate the usability of \sys{}.
We aimed to see if \sys{} could provide users with an effective digital resource curation method in an ecologically valid setting. 

\subsection{Recruitment}


We recruited 13 participants (7 male, 5 female, and 1 non-binary) from the authors' university mailing list and \ADD{the Slack workspace of a research center at the authors' university}.\marking{5}
Participants had an average age of 26.4, ranging from 19 to 37.
The participants were mostly graduate students, as well as two undergraduate students, one software engineer, and one data analyst. 
We consider university students to be an eligible target population that shares the challenges that knowledge workers encounter as they use computers as essential devices for research and study; the authors' university requires all university students to have their own personal computing devices when they are admitted.  
We limited participants to those who use Apple's macOS operating system for their work or studies on a daily basis. 
The average self-assessed daily computer usage time was 9.4 hours ($\sigma$ = 3.4, maximum = 16, minimum = 6).
The results of the pre-study survey indicated that all participants reported their digital literacy~\cite{doi:10.1080/13614533.2015.1137466} as ``above average'' (7/13) or ``excellent'' (6/13) on a five-point scale (``poor,'' ``below average,'' ``average,'' ``above average,'' and ``excellent'').

\subsection{Field Study}
The entire study was done remotely. 
Initially, we held a remote meeting in which the first author demonstrated how to use \sys{}; guided participants to install \sys{} and required services (e.g., MongoDB, Node.js) on their computers; explained details and requirements concerning the user study, and asked them to fill out a demographic survey. 
After the meeting, we instructed participants to use their computers for their daily computing tasks as usual for a week.
We asked them to use \sys{} during that time to curate digital resources and take a few screenshots each day.
\ADD{Participants were not otherwise instructed to perform specific tasks on their computers.}\marking{5}
After one week of using \sys{}, we conducted a semi-structured interview with participants to ask them questions about their experience of using \sys{}.
Most of the questions concerned how participants used \sys{}, their motivation for taking screenshots, how they liked or disliked the system, and what concerns or suggested improvements they had for the system. For a full list of interview questions, see the appendix. 
The initial meeting and exit interview took about 1.5 hours, and participants were compensated with e-gift cards worth \$40. 

\subsection{Data Collection and Analysis}

We calculated descriptive statistics on participants' use of \sys{} to show their levels of active usage. 
We developed a logging tool to track how many times participants created screenshots and reopened resources from \sys{}. 
We collected metadata and screenshots only if participants agreed to share this data, as their screenshots could have included personal information. 
All participants shared complete metadata; \ADD{most of the participants shared the entire set of screenshots, while some shared a selection of screenshots.} 
\ADD{Participants who declined to share some of their screenshots were concerned about the security of personal information captured in the screenshots.}
\ADD{The first author performed a thematic analysis and generated initial codes from the collected interview data. 
Then, both authors discussed the codes and corresponding quotes iteratively until they fully agreed on the final result by revising, merging, and deleting the initial codes.}\marking{5}

\section{Results}

\subsection{Descriptive Statistics for \sys{} Usage}\marking{6}
During the one-week long field study, participants used \sys{} to capture \ADD{212 screenshots in total, with }an average of 16.3 screenshots per user ($\sigma$ = 6.2, maximum = 32, minimum = 6). 
\ADD{Out of all the screenshots, 62.3\% captured a single visible application window.}
The average number of visible windows per screenshot was 1.55 ($\sigma$ = 0.60, \ADD{maximum = 5}). 
\ADD{Taking invisible windows into account, each participant selected and saved 3.13 ($\sigma$ = 1.94) windows on average per curation event.
This result shows that users added invisible windows alongside their visible windows.}
Most participants took both types of screenshot, with the exception of four who took only full-screen screenshots; 64.2\% of all captured screenshots were full-screen screenshots, and the rest were selected-area screenshots. 
\ADD{
Recorded computer activities and digital resources varied across individuals, from simple web surfing activity to complicated programming tasks.}
Participants' captures covered a wide range of applications, the majority being web browsers (63.5\%), followed by office software (7.6\%), PDF viewers (4.9\%), file managers (4.9\%), email applications (3.5\%), programming editors (2.7\%), and others.
\ADD{While the majority of screenshots featured web browsers, 26.2\% of screenshots contained only web browsers.
This result shows that existing bookmark functionality, which only covers web pages, would have been able to cover participants' needs to bundle various resources.}
On average, users opened applications from screenshots 0.78 times per day. 
While this is below the average number of screenshots taken per day, it could have been the case that the need to reopen some resources did not occur during the study period. 
Also, we later found out from our interviews that many participants simply took screenshots as visual reminders. 
Overall, the descriptive statistics indicate that participants engaged with \sys{} during the study, which allowed them to reflect on the user experience during the exit interviews. 


\subsection{Recognition of Work at a Glance}

Participants found that using screenshots was effective in recognizing what a task was about at a glance. 
Many participants (9/13) explicitly mentioned how easy it was to recognize the subject of a screenshot. 
Here is one participant's comment that captures their perception of screenshot-based bookmarks: 

\begin{quote}[P3]
``If I'm just looking at this [Collection View] for a second, like, you know, images grab your eye, and you might actually remember something and be like, `Oh, what is that?' Like \ldots So I think I'm more likely to remember. I think pictures are good because they jog your memory, like, pretty instantly. And you don't really have to write and they just catch your attention.'' 
\end{quote}

\noindent In particular, many participants (8/13) specifically mentioned how useful the Collection View was, perhaps even more often than the Detailed View, because they were able to quickly grasp their portfolio of resources at a glance.  


\begin{quote}[P2]
(\textit{Responding to a follow-up question asking if user-supplied text, such as titles and descriptions, was helpful}) ``Um, for me, the screenshot was enough, because in the Collection View, I can already see every picture and just by, like, seeing that quick image of it, I know what it is. I did, like, put descriptions for all of mine, just, like, as an example to, like, say, what the specific part is about. But typically, I just look at the picture. I don't really look at the description part or the title part.''
\end{quote}
   


\noindent We realize that one of the critical values of \sys{} is simply having a central repository for all screenshots. 
 
\begin{figure*}[t]
  \centering
  \includegraphics[width=\textwidth]{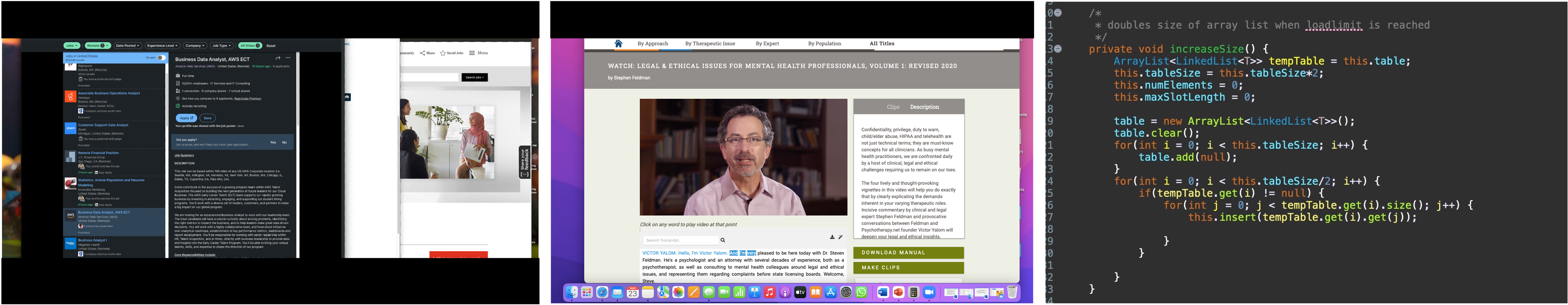}
  \caption{Example screenshots from participants. Some parts of the images were redacted to protect participants' identities. Left: P10 captured multiple windows in the context of a job search for later task resumption. Middle: P1 captured a video that they wanted to revisit later. Right: P3 took a snapshot of their code for temporary version control.}
  \label{fig:code}
\end{figure*}
\subsection{Glanceable Bookmarks and To-do List}

In terms of the motivation behind taking screenshots, \sys{} supported various participant needs. 
Most participants (11/13) perceived the benefit of bookmark functionality; that is, being able to open resources from a screenshot.
Of the 13 participants interviewed, 10 mentioned that \sys{} served as a cross-application bookmark manager that they used to curate multiple digital resources into bundles. 
Other participants perceived a screenshot as a good way to keep track of tasks to complete (9/13). The following comments from P8 and P7 highlight how the Collection View helped them review their work. 

\begin{quote}[P8]
``To keep a record for myself. And then, if later I want to reflect back on what I was doing on that day, I may be able to look at screenshots.''
\end{quote}
\begin{quote}[P7]
``I think \sys{} is very good for reviewing the screenshots, because most oftentimes, when I screenshot something, I have to put it somewhere. And because it's in one place, I can search in it. Yeah, just wonderful. And I can have a title, and then searching the titles is very useful.''
\end{quote}

\noindent Some participants (3/13) used screenshots to increase their comfort with closing multiple applications. 

In one interesting use case, a participant captured a screenshot of their code editor as a way to capture a version of their code before revising it (Figure~\ref{fig:code}-Right).
While the earlier version of the code would not have been available when opening the updated file from the screenshot, the screen captured one version that he believed it was stable. 
Interestingly, the user was actually able to select and copy code text from the image using their web browser's built-in optical character recognition (OCR) functionality\footnote{\url{https://support.apple.com/guide/safari/interact-with-text-in-a-picture-ibrw20183ad7/mac}}, which fulfilled the purpose that the user intended. 
\subsection{Specificity and Transiency}

One interesting trend we observed from our visual inspection of shared screenshots was that screenshots served more as short-term bookmarks that were useful to resume tasks or remind users of tasks to complete. 
For example, one participant (P3) mentioned that they had a bookmark for the front page of a system in a web browser; the screenshot-based bookmark was linked to a specific page within the website. 
The example the participant gave us was a learning management system (e.g., Canvas).
Using a traditional bookmark, they bookmarked the front page and had to then navigate the course website in multiple steps if they wanted to resume a grading task: expand the menu, select a course, choose a menu item, select an assignment, open a grader, and find the student they needed to resume grading from. 
In contrast, the screenshot bookmark they created in \sys{} was a link to a specific student's homework submission that they needed to resume grading from, resembling the original application of physical bookmarks: marking progress within a book, rather than merely marking the entire book. 
Similar examples of specific bookmarks are also available in Figure~\ref{fig:code}-Left and -Middle). 
P1, who took the screenshot in Figure~\ref{fig:code}-Middle, stated that they bookmarked the website before the study and ordinarily had to navigate from the front page to get to the video. 
However, using \sys{}, they bookmarked a page for a specific video they had been watching. 
While it is not impossible to bookmark a specific page with a web browser's bookmark functionality, it seems that the temporally organized collection view with visual cues encouraged participants to create transient bookmarks for task resumption, as opposed to bookmarking gateways to specific resources. 
  




\subsection{Concerns and Suggestions}

We asked if participants had any concerns or suggestions regarding \sys{}.
While we explicitly mentioned that the system was a fully local application that does not communicate with other services, some participants (6/13) were concerned about personal information appearing in screenshots and wondered how much they could trust an app developer in practice. 
There were many suggestions regarding the system. The most popular suggestion (6/13) was to add functionality for organizing screenshots. 
Some participants suggested more traditional approaches, such as a hierarchical folder structure, even though we intentionally moved away from such approaches to minimize the effort expended to maintain such a hierarchy. 
Others wanted more ways to filter out subsets of screenshots quickly, with tags and more powerful search functions. 
One participant (P3) proposed that it would be excellent if \sys{} could reproduce the window layout as well when restoring windows, which is feasible. 
Another interesting suggestion was to store all the tabs in a web browser window. 
We specifically avoided this functionality because it could capture too many tabs and overwhelm users with a long list of tabs in the Capture View.  




\section{Discussion and Future Work}

The result of our one-week field study indicates that the participants were able to effectively curate resources for various purposes, from task resumption to bookmarking, for potential revisiting. 
In general, the participants were able to reconstruct contexts from screenshots, and the Collection View served as a glanceable curated list for keeping track of tasks and resources.

One design implication we found is that the Collection View design nudged participants to use screenshot-bookmarks for short-term reminders as well as long-term resource management.  
Therefore, screenshots of heterogeneous types can coexist in the collection: some are temporary and others are permanent. 
Reminder bookmarks will be transient, such as a to-do item that is meaningful only until its completion. 
Therefore, some effort may be required of users to delete or archive completed screenshots or categorize them during the capture process. 
Making it easy to archive a screenshot, similar to ticking a checkbox in a to-do list, could be another way to support reminder-like usage of screenshot-based bookmarks. 
In addition, automatically archiving screenshots based on type or making certain screenshots ephemeral could be useful to maintain a reasonable scale. 


We noticed that screenshot-based bookmarks encouraged users to create specific, transient screenshots that were easily recognizable.
This tendency means that screenshot-based bookmarks remain viable at a much larger scale than browser bookmarks.
Consequently, organization methods, such as filtering by time and type, searching, tags, and sorting, will be crucial in practice; manual curation (such as creating a folder hierarchy) should be avoided, as it is not scalable.
Additionally, extracting semantic information from a screenshot and its metadata to automatically suggest a semantically meaningful title would minimize the effort expended on user-supplied information and make screenshots more searchable.

Given that the vast majority of captures contained a web browser, the additional information dimension of multiple tabs poses a technical challenge. 
Storing all of the tabs or asking users to choose relevant tabs from the entire set would undermine the value in the minimal effort needed to create screenshot-based bookmarks. 
However, capturing only one tab per screenshot would be unrealistic. 
Therefore, we are in need of an intelligent algorithm that can evaluate the relevance of inactive tabs to the visible tab based on proximity, how and when a new tab was opened, and transitions between tabs. 
Identifying interdependency between multiple tabs by usage pattern will be a promising research topic for reducing manual organization in tab bundling approaches~\cite{10.1145/3173574.3173825,10.1145/3472749.3474777}. 

\ADD{
We foresee future works in using screenshots as a medium to support activity-based computing~\cite{bardram2005activity}. 
Screenshots can be used to support mobility by allowing users to distribute resources across devices, such as a home computer, computers at work, and tablets. 
In addition, screenshots can serve as a medium to share resources needed for both synchronous and asynchronous collaboration (e.g., sharing URLs and files as a bundle by sharing a screenshot). 
}

 
Lastly, we envision that intelligent screenshot recommendations may alleviate the burden of resource curation. 
Productivity tracking tools such as RescueTime and ScreenTrack~\cite{hu2020screentrack} can continuously monitor computer activities and provide metadata that can be used to recommend a screenshot that a user may want to curate. 
The importance of a given state of a computer screen can be evaluated in retrospect with various metadata---visible windows, window layouts, and transition patterns---to provide users recommended screenshots for curation and place them in personalized layouts for their computing environments.

\section{Conclusions}
\ADD{We presented \sys{}, a novel application that can help knowledge workers curate cross-application resources by taking screenshots. 
There are multiple benefits that come from using visual cues along with other symbolic information, including enhanced recognizability and the ability to bundle multiple windows into one handle. 
We developed a bit-masking algorithm to identify the visibility of captured windows. 
Additionally, the results of the user study indicate that participants gave positive feedback on using Scrapbook and were able to effectively curate digital resources for various purposes. 
We anticipate that our screenshot-based productivity tool will offer users additional ways to access and organize digital resources and cope with information overload.}\marking{7} 

\begin{acks}
We thank all the reviewers for their thorough feedback. We are also grateful to all user study participants for their participation and sharing their screenshot data.  
\end{acks}

\bibliographystyle{ACM-Reference-Format}
\bibliography{sample-sigconf}


\begin{thebibliography}{49}


\ifx \showCODEN    \undefined \def \showCODEN     #1{\unskip}     \fi
\ifx \showDOI      \undefined \def \showDOI       #1{#1}\fi
\ifx \showISBNx    \undefined \def \showISBNx     #1{\unskip}     \fi
\ifx \showISBNxiii \undefined \def \showISBNxiii  #1{\unskip}     \fi
\ifx \showISSN     \undefined \def \showISSN      #1{\unskip}     \fi
\ifx \showLCCN     \undefined \def \showLCCN      #1{\unskip}     \fi
\ifx \shownote     \undefined \def \shownote      #1{#1}          \fi
\ifx \showarticletitle \undefined \def \showarticletitle #1{#1}   \fi
\ifx \showURL      \undefined \def \showURL       {\relax}        \fi
\providecommand\bibfield[2]{#2}
\providecommand\bibinfo[2]{#2}
\providecommand\natexlab[1]{#1}
\providecommand\showeprint[2][]{arXiv:#2}

\bibitem[Abrams et~al\mbox{.}(1998)]%
        {10.1145/274644.274651}
\bibfield{author}{\bibinfo{person}{David Abrams}, \bibinfo{person}{Ron
  Baecker}, {and} \bibinfo{person}{Mark Chignell}.}
  \bibinfo{year}{1998}\natexlab{}.
\newblock \showarticletitle{Information Archiving with Bookmarks: Personal Web
  Space Construction and Organization}. In
  \bibinfo{booktitle}{\emph{Proceedings of the SIGCHI Conference on Human
  Factors in Computing Systems}} (Los Angeles, California, USA)
  \emph{(\bibinfo{series}{CHI '98})}. \bibinfo{publisher}{ACM
  Press/Addison-Wesley Publishing Co.}, \bibinfo{address}{USA},
  \bibinfo{pages}{41–48}.
\newblock
\showISBNx{0201309874}
\urldef\tempurl%
\url{https://doi.org/10.1145/274644.274651}
\showDOI{\tempurl}


\bibitem[Bannon et~al\mbox{.}(1983)]%
        {bannon1983evaluation}
\bibfield{author}{\bibinfo{person}{Liam Bannon}, \bibinfo{person}{Allen
  Cypher}, \bibinfo{person}{Steven Greenspan}, {and} \bibinfo{person}{Melissa~L
  Monty}.} \bibinfo{year}{1983}\natexlab{}.
\newblock \showarticletitle{Evaluation and analysis of users' activity
  organization}. In \bibinfo{booktitle}{\emph{Proceedings of the SIGCHI
  conference on Human Factors in Computing Systems}}. \bibinfo{pages}{54--57}.
\newblock


\bibitem[Bardram(2005)]%
        {bardram2005activity}
\bibfield{author}{\bibinfo{person}{Jakob~E Bardram}.}
  \bibinfo{year}{2005}\natexlab{}.
\newblock \showarticletitle{Activity-based computing: support for mobility and
  collaboration in ubiquitous computing}.
\newblock \bibinfo{journal}{\emph{Personal and Ubiquitous Computing}}
  \bibinfo{volume}{9}, \bibinfo{number}{5} (\bibinfo{year}{2005}),
  \bibinfo{pages}{312--322}.
\newblock


\bibitem[Bernstein et~al\mbox{.}(2008c)]%
        {bernstein2008information}
\bibfield{author}{\bibinfo{person}{Michael Bernstein}, \bibinfo{person}{Max
  Van~Kleek}, \bibinfo{person}{David Karger}, {and} \bibinfo{person}{MC
  Schraefel}.} \bibinfo{year}{2008}\natexlab{c}.
\newblock \showarticletitle{Information scraps: How and why information eludes
  our personal information management tools}.
\newblock \bibinfo{journal}{\emph{ACM Transactions on Information Systems
  (TOIS)}} \bibinfo{volume}{26}, \bibinfo{number}{4} (\bibinfo{year}{2008}),
  \bibinfo{pages}{1--46}.
\newblock


\bibitem[Bernstein et~al\mbox{.}(2008a)]%
        {Bernstein:2008:TEF:1449715.1449753}
\bibfield{author}{\bibinfo{person}{Michael~S. Bernstein}, \bibinfo{person}{Jeff
  Shrager}, {and} \bibinfo{person}{Terry Winograd}.}
  \bibinfo{year}{2008}\natexlab{a}.
\newblock \showarticletitle{Taskpos{\'e}: Exploring Fluid Boundaries in an
  Associative Window Visualization}. In \bibinfo{booktitle}{\emph{Proceedings
  of the 21st Annual ACM Symposium on User Interface Software and Technology}}
  (Monterey, CA, USA) \emph{(\bibinfo{series}{UIST '08})}.
  \bibinfo{publisher}{ACM}, \bibinfo{address}{New York, NY, USA},
  \bibinfo{pages}{231--234}.
\newblock
\showISBNx{978-1-59593-975-3}
\urldef\tempurl%
\url{https://doi.org/10.1145/1449715.1449753}
\showDOI{\tempurl}


\bibitem[Bernstein et~al\mbox{.}(2008b)]%
        {bernstein2008taskpose}
\bibfield{author}{\bibinfo{person}{Michael~S Bernstein}, \bibinfo{person}{Jeff
  Shrager}, {and} \bibinfo{person}{Terry Winograd}.}
  \bibinfo{year}{2008}\natexlab{b}.
\newblock \showarticletitle{Taskpos{\'e}: exploring fluid boundaries in an
  associative window visualization}. In \bibinfo{booktitle}{\emph{Proceedings
  of the 21st annual ACM symposium on User interface software and technology}}.
  \bibinfo{pages}{231--234}.
\newblock


\bibitem[Bernstein et~al\mbox{.}(2007)]%
        {bernstein2007management}
\bibfield{author}{\bibinfo{person}{Michael~S Bernstein}, \bibinfo{person}{Max
  Van~Kleek}, \bibinfo{person}{MC Schraefel}, {and} \bibinfo{person}{David~R
  Karger}.} \bibinfo{year}{2007}\natexlab{}.
\newblock \showarticletitle{Management of personal information scraps}. In
  \bibinfo{booktitle}{\emph{CHI'07 Extended Abstracts on Human Factors in
  Computing Systems}}. \bibinfo{pages}{2285--2290}.
\newblock


\bibitem[Brady et~al\mbox{.}(2008)]%
        {brady2008visual}
\bibfield{author}{\bibinfo{person}{Timothy~F Brady}, \bibinfo{person}{Talia
  Konkle}, \bibinfo{person}{George~A Alvarez}, {and} \bibinfo{person}{Aude
  Oliva}.} \bibinfo{year}{2008}\natexlab{}.
\newblock \showarticletitle{Visual long-term memory has a massive storage
  capacity for object details}.
\newblock \bibinfo{journal}{\emph{Proceedings of the National Academy of
  Sciences}} \bibinfo{volume}{105}, \bibinfo{number}{38}
  (\bibinfo{year}{2008}), \bibinfo{pages}{14325--14329}.
\newblock


\bibitem[Chang et~al\mbox{.}(2021)]%
        {10.1145/3472749.3474777}
\bibfield{author}{\bibinfo{person}{Joseph~Chee Chang},
  \bibinfo{person}{Yongsung Kim}, \bibinfo{person}{Victor Miller},
  \bibinfo{person}{Michael~Xieyang Liu}, \bibinfo{person}{Brad~A Myers}, {and}
  \bibinfo{person}{Aniket Kittur}.} \bibinfo{year}{2021}\natexlab{}.
\newblock \showarticletitle{Tabs.Do: Task-Centric Browser Tab Management}. In
  \bibinfo{booktitle}{\emph{The 34th Annual ACM Symposium on User Interface
  Software and Technology}} (Virtual Event, USA) \emph{(\bibinfo{series}{UIST
  '21})}. \bibinfo{publisher}{Association for Computing Machinery},
  \bibinfo{address}{New York, NY, USA}, \bibinfo{pages}{663–676}.
\newblock
\showISBNx{9781450386357}
\urldef\tempurl%
\url{https://doi.org/10.1145/3472749.3474777}
\showDOI{\tempurl}


\bibitem[Cockburn and McKenzie(2001)]%
        {cockburn2001web}
\bibfield{author}{\bibinfo{person}{Andy Cockburn} {and} \bibinfo{person}{Bruce
  McKenzie}.} \bibinfo{year}{2001}\natexlab{}.
\newblock \showarticletitle{What do web users do? An empirical analysis of web
  use}.
\newblock \bibinfo{journal}{\emph{International Journal of human-computer
  studies}} \bibinfo{volume}{54}, \bibinfo{number}{6} (\bibinfo{year}{2001}),
  \bibinfo{pages}{903--922}.
\newblock


\bibitem[Czerwinski and Horvitz(2002)]%
        {czerwinski2002investigation}
\bibfield{author}{\bibinfo{person}{Mary Czerwinski} {and} \bibinfo{person}{Eric
  Horvitz}.} \bibinfo{year}{2002}\natexlab{}.
\newblock \showarticletitle{An investigation of memory for daily computing
  events}.
\newblock In \bibinfo{booktitle}{\emph{People and computers XVI-memorable yet
  invisible}}. \bibinfo{publisher}{Springer}, \bibinfo{pages}{229--245}.
\newblock


\bibitem[Czerwinski et~al\mbox{.}(2004)]%
        {czerwinski2004diary}
\bibfield{author}{\bibinfo{person}{Mary Czerwinski}, \bibinfo{person}{Eric
  Horvitz}, {and} \bibinfo{person}{Susan Wilhite}.}
  \bibinfo{year}{2004}\natexlab{}.
\newblock \showarticletitle{A diary study of task switching and interruptions}.
  In \bibinfo{booktitle}{\emph{Proceedings of the SIGCHI conference on Human
  factors in computing systems}}. \bibinfo{pages}{175--182}.
\newblock


\bibitem[Docs(2001)]%
        {Thornburg01}
\bibfield{author}{\bibinfo{person}{MDN~Web Docs}.}
  \bibinfo{year}{2001}\natexlab{}.
\newblock \bibinfo{booktitle}{\emph{Masonry layout CSS: Cascading style sheets:
  MDN}}.
\newblock
\urldef\tempurl%
\url{https://developer.mozilla.org/en-US/docs/Web/CSS/CSS_Grid_Layout/Masonry_Layout}
\showURL{%
Retrieved April 7, 2022 from \tempurl}


\bibitem[Dragunov et~al\mbox{.}(2005a)]%
        {dragunov2005tasktracer}
\bibfield{author}{\bibinfo{person}{Anton~N Dragunov}, \bibinfo{person}{Thomas~G
  Dietterich}, \bibinfo{person}{Kevin Johnsrude}, \bibinfo{person}{Matthew
  McLaughlin}, \bibinfo{person}{Lida Li}, {and} \bibinfo{person}{Jonathan~L
  Herlocker}.} \bibinfo{year}{2005}\natexlab{a}.
\newblock \showarticletitle{TaskTracer: a desktop environment to support
  multi-tasking knowledge workers}. In \bibinfo{booktitle}{\emph{Proceedings of
  the 10th international conference on Intelligent user interfaces}}.
  \bibinfo{pages}{75--82}.
\newblock


\bibitem[Dragunov et~al\mbox{.}(2005b)]%
        {Dragunov:2005:TDE:1040830.1040855_tasktracer}
\bibfield{author}{\bibinfo{person}{Anton~N. Dragunov},
  \bibinfo{person}{Thomas~G. Dietterich}, \bibinfo{person}{Kevin Johnsrude},
  \bibinfo{person}{Matthew McLaughlin}, \bibinfo{person}{Lida Li}, {and}
  \bibinfo{person}{Jonathan~L. Herlocker}.} \bibinfo{year}{2005}\natexlab{b}.
\newblock \showarticletitle{TaskTracer: A Desktop Environment to Support
  Multi-tasking Knowledge Workers}. In \bibinfo{booktitle}{\emph{Proceedings of
  the 10th International Conference on Intelligent User Interfaces}} (San
  Diego, California, USA) \emph{(\bibinfo{series}{IUI '05})}.
  \bibinfo{publisher}{ACM}, \bibinfo{address}{New York, NY, USA},
  \bibinfo{pages}{75--82}.
\newblock
\showISBNx{1-58113-894-6}
\urldef\tempurl%
\url{https://doi.org/10.1145/1040830.1040855}
\showDOI{\tempurl}


\bibitem[Finley et~al\mbox{.}(2018)]%
        {Finley2018}
\bibfield{author}{\bibinfo{person}{Jason~R. Finley}, \bibinfo{person}{Farah
  Naaz}, {and} \bibinfo{person}{Francine~W. Goh}.}
  \bibinfo{year}{2018}\natexlab{}.
\newblock \bibinfo{booktitle}{\emph{Results: Behaviors and Experiences with
  Internal and External Memory}}.
\newblock \bibinfo{publisher}{Springer International Publishing},
  \bibinfo{address}{Cham}, \bibinfo{pages}{25--48}.
\newblock
\showISBNx{978-3-319-99169-6}
\urldef\tempurl%
\url{https://doi.org/10.1007/978-3-319-99169-6_3}
\showDOI{\tempurl}


\bibitem[Gonz{\'a}lez and Mark(2004)]%
        {gonzalez2004constant}
\bibfield{author}{\bibinfo{person}{Victor~M Gonz{\'a}lez} {and}
  \bibinfo{person}{Gloria Mark}.} \bibinfo{year}{2004}\natexlab{}.
\newblock \showarticletitle{" Constant, constant, multi-tasking craziness"
  managing multiple working spheres}. In \bibinfo{booktitle}{\emph{Proceedings
  of the SIGCHI conference on Human factors in computing systems}}.
  \bibinfo{pages}{113--120}.
\newblock


\bibitem[Grevet et~al\mbox{.}(2014)]%
        {10.1145/2556288.2557013}
\bibfield{author}{\bibinfo{person}{Catherine Grevet}, \bibinfo{person}{David
  Choi}, \bibinfo{person}{Debra Kumar}, {and} \bibinfo{person}{Eric Gilbert}.}
  \bibinfo{year}{2014}\natexlab{}.
\newblock \bibinfo{booktitle}{\emph{Overload is Overloaded: Email in the Age of
  Gmail}}.
\newblock \bibinfo{publisher}{Association for Computing Machinery},
  \bibinfo{address}{New York, NY, USA}, \bibinfo{pages}{793–802}.
\newblock
\showISBNx{9781450324731}
\urldef\tempurl%
\url{https://doi.org/10.1145/2556288.2557013}
\showURL{%
\tempurl}


\bibitem[Hahn et~al\mbox{.}(2018)]%
        {10.1145/3173574.3173825}
\bibfield{author}{\bibinfo{person}{Nathan Hahn}, \bibinfo{person}{Joseph~Chee
  Chang}, {and} \bibinfo{person}{Aniket Kittur}.}
  \bibinfo{year}{2018}\natexlab{}.
\newblock \showarticletitle{Bento Browser: Complex Mobile Search Without Tabs}.
  In \bibinfo{booktitle}{\emph{Proceedings of the 2018 CHI Conference on Human
  Factors in Computing Systems}} (Montreal QC, Canada)
  \emph{(\bibinfo{series}{CHI '18})}. \bibinfo{publisher}{Association for
  Computing Machinery}, \bibinfo{address}{New York, NY, USA},
  \bibinfo{pages}{1–12}.
\newblock
\showISBNx{9781450356206}
\urldef\tempurl%
\url{https://doi.org/10.1145/3173574.3173825}
\showDOI{\tempurl}


\bibitem[Hailpern et~al\mbox{.}(2011)]%
        {Hailpern:2011:YIR:1978942.1979165}
\bibfield{author}{\bibinfo{person}{Joshua Hailpern}, \bibinfo{person}{Nicholas
  Jitkoff}, \bibinfo{person}{Andrew Warr}, \bibinfo{person}{Karrie Karahalios},
  \bibinfo{person}{Robert Sesek}, {and} \bibinfo{person}{Nik Shkrob}.}
  \bibinfo{year}{2011}\natexlab{}.
\newblock \showarticletitle{YouPivot: Improving Recall with Contextual Search}.
  In \bibinfo{booktitle}{\emph{Proceedings of the SIGCHI Conference on Human
  Factors in Computing Systems}} (Vancouver, BC, Canada)
  \emph{(\bibinfo{series}{CHI '11})}. \bibinfo{publisher}{ACM},
  \bibinfo{address}{New York, NY, USA}, \bibinfo{pages}{1521--1530}.
\newblock
\showISBNx{978-1-4503-0228-9}
\urldef\tempurl%
\url{https://doi.org/10.1145/1978942.1979165}
\showDOI{\tempurl}


\bibitem[Hu and Lee(2020)]%
        {hu2020screentrack}
\bibfield{author}{\bibinfo{person}{Donghan Hu} {and} \bibinfo{person}{Sang~Won
  Lee}.} \bibinfo{year}{2020}\natexlab{}.
\newblock \showarticletitle{ScreenTrack: Using a Visual History of a Computer
  Screen to Retrieve Documents and Web Pages}. In
  \bibinfo{booktitle}{\emph{Proceedings of the 2020 CHI Conference on Human
  Factors in Computing Systems}}. \bibinfo{pages}{1--13}.
\newblock


\bibitem[Iqbal and Horvitz(2007)]%
        {iqbal2007disruption}
\bibfield{author}{\bibinfo{person}{Shamsi~T Iqbal} {and} \bibinfo{person}{Eric
  Horvitz}.} \bibinfo{year}{2007}\natexlab{}.
\newblock \showarticletitle{Disruption and recovery of computing tasks: field
  study, analysis, and directions}. In \bibinfo{booktitle}{\emph{Proceedings of
  the SIGCHI conference on Human factors in computing systems}}.
  \bibinfo{pages}{677--686}.
\newblock


\bibitem[Jacobson(2012)]%
        {jacobson2012information}
\bibfield{author}{\bibinfo{person}{Jenna Jacobson}.}
  \bibinfo{year}{2012}\natexlab{}.
\newblock \showarticletitle{Information curation}.
\newblock \bibinfo{journal}{\emph{Proceedings of the American Society for
  Information Science and Technology}} \bibinfo{volume}{49},
  \bibinfo{number}{1} (\bibinfo{year}{2012}), \bibinfo{pages}{1--2}.
\newblock


\bibitem[Ko et~al\mbox{.}(2005)]%
        {ko2005eliciting}
\bibfield{author}{\bibinfo{person}{Amy~J Ko}, \bibinfo{person}{Htet Aung},
  {and} \bibinfo{person}{Brad~A Myers}.} \bibinfo{year}{2005}\natexlab{}.
\newblock \showarticletitle{Eliciting design requirements for
  maintenance-oriented ides: a detailed study of corrective and perfective
  maintenance tasks}. In \bibinfo{booktitle}{\emph{Proceedings of the 27th
  international conference on Software engineering}}.
  \bibinfo{pages}{126--135}.
\newblock


\bibitem[Mark et~al\mbox{.}(2015)]%
        {mark2015focused}
\bibfield{author}{\bibinfo{person}{Gloria Mark}, \bibinfo{person}{Shamsi
  Iqbal}, \bibinfo{person}{Mary Czerwinski}, {and} \bibinfo{person}{Paul
  Johns}.} \bibinfo{year}{2015}\natexlab{}.
\newblock \showarticletitle{Focused, aroused, but so distractible: Temporal
  perspectives on multitasking and communications}. In
  \bibinfo{booktitle}{\emph{Proceedings of the 18th ACM Conference on Computer
  Supported Cooperative Work \& Social Computing}}. \bibinfo{pages}{903--916}.
\newblock


\bibitem[Morris et~al\mbox{.}(2008)]%
        {10.1145/1357054.1357242}
\bibfield{author}{\bibinfo{person}{Dan Morris}, \bibinfo{person}{Meredith
  Ringel~Morris}, {and} \bibinfo{person}{Gina Venolia}.}
  \bibinfo{year}{2008}\natexlab{}.
\newblock \showarticletitle{SearchBar: A Search-Centric Web History for Task
  Resumption and Information Re-Finding}. In
  \bibinfo{booktitle}{\emph{Proceedings of the SIGCHI Conference on Human
  Factors in Computing Systems}} (Florence, Italy) \emph{(\bibinfo{series}{CHI
  '08})}. \bibinfo{publisher}{Association for Computing Machinery},
  \bibinfo{address}{New York, NY, USA}, \bibinfo{pages}{1207–1216}.
\newblock
\showISBNx{9781605580111}
\urldef\tempurl%
\url{https://doi.org/10.1145/1357054.1357242}
\showDOI{\tempurl}


\bibitem[Morris and Horvitz(2007)]%
        {10.1145/1294211.1294215}
\bibfield{author}{\bibinfo{person}{Meredith~Ringel Morris} {and}
  \bibinfo{person}{Eric Horvitz}.} \bibinfo{year}{2007}\natexlab{}.
\newblock \showarticletitle{SearchTogether: An Interface for Collaborative Web
  Search}. In \bibinfo{booktitle}{\emph{Proceedings of the 20th Annual ACM
  Symposium on User Interface Software and Technology}} (Newport, Rhode Island,
  USA) \emph{(\bibinfo{series}{UIST '07})}. \bibinfo{publisher}{Association for
  Computing Machinery}, \bibinfo{address}{New York, NY, USA},
  \bibinfo{pages}{3–12}.
\newblock
\showISBNx{9781595936790}
\urldef\tempurl%
\url{https://doi.org/10.1145/1294211.1294215}
\showDOI{\tempurl}


\bibitem[Oleksik et~al\mbox{.}(2009)]%
        {oleksik2009lightweight}
\bibfield{author}{\bibinfo{person}{Gerard Oleksik}, \bibinfo{person}{Max~L
  Wilson}, \bibinfo{person}{Craig Tashman}, \bibinfo{person}{Eduarda
  Mendes~Rodrigues}, \bibinfo{person}{Gabriella Kazai}, \bibinfo{person}{Gavin
  Smyth}, \bibinfo{person}{Natasa Milic-Frayling}, {and}
  \bibinfo{person}{Rachel Jones}.} \bibinfo{year}{2009}\natexlab{}.
\newblock \showarticletitle{Lightweight tagging expands information and
  activity management practices}. In \bibinfo{booktitle}{\emph{Proceedings of
  the SIGCHI Conference on Human Factors in Computing Systems}}.
  \bibinfo{pages}{279--288}.
\newblock


\bibitem[Parnin and DeLine(2010)]%
        {parnin2010evaluating}
\bibfield{author}{\bibinfo{person}{Chris Parnin} {and} \bibinfo{person}{Robert
  DeLine}.} \bibinfo{year}{2010}\natexlab{}.
\newblock \showarticletitle{Evaluating cues for resuming interrupted
  programming tasks}. In \bibinfo{booktitle}{\emph{Proceedings of the SIGCHI
  conference on human factors in computing systems}}. \bibinfo{pages}{93--102}.
\newblock


\bibitem[Parnin and Rugaber(2011)]%
        {parnin2011resumption}
\bibfield{author}{\bibinfo{person}{Chris Parnin} {and} \bibinfo{person}{Spencer
  Rugaber}.} \bibinfo{year}{2011}\natexlab{}.
\newblock \showarticletitle{Resumption strategies for interrupted programming
  tasks}.
\newblock \bibinfo{journal}{\emph{Software Quality Journal}}
  \bibinfo{volume}{19}, \bibinfo{number}{1} (\bibinfo{year}{2011}),
  \bibinfo{pages}{5--34}.
\newblock


\bibitem[Robertson et~al\mbox{.}(2004)]%
        {ScalableFabric}
\bibfield{author}{\bibinfo{person}{George Robertson}, \bibinfo{person}{Eric
  Horvitz}, \bibinfo{person}{Mary Czerwinski}, \bibinfo{person}{Patrick
  Baudisch}, \bibinfo{person}{Dugald~Ralph Hutchings}, \bibinfo{person}{Brian
  Meyers}, \bibinfo{person}{Daniel Robbins}, {and} \bibinfo{person}{Greg
  Smith}.} \bibinfo{year}{2004}\natexlab{}.
\newblock \showarticletitle{Scalable Fabric: Flexible Task Management}. In
  \bibinfo{booktitle}{\emph{Proceedings of the Working Conference on Advanced
  Visual Interfaces}} (Gallipoli, Italy) \emph{(\bibinfo{series}{AVI '04})}.
  \bibinfo{publisher}{Association for Computing Machinery},
  \bibinfo{address}{New York, NY, USA}, \bibinfo{pages}{85–89}.
\newblock
\showISBNx{1581138679}
\urldef\tempurl%
\url{https://doi.org/10.1145/989863.989874}
\showDOI{\tempurl}


\bibitem[Rule et~al\mbox{.}(2015)]%
        {rule2015restoring}
\bibfield{author}{\bibinfo{person}{Adam Rule}, \bibinfo{person}{Aur{\'e}lien
  Tabard}, \bibinfo{person}{Karen Boyd}, {and} \bibinfo{person}{Jim Hollan}.}
  \bibinfo{year}{2015}\natexlab{}.
\newblock \showarticletitle{Restoring the context of interrupted work with
  desktop thumbnails}.
\newblock


\bibitem[Rule et~al\mbox{.}(2017)]%
        {rule2017using}
\bibfield{author}{\bibinfo{person}{Adam Rule}, \bibinfo{person}{Aur{\'e}lien
  Tabard}, {and} \bibinfo{person}{Jim Hollan}.}
  \bibinfo{year}{2017}\natexlab{}.
\newblock \showarticletitle{Using visual histories to reconstruct the mental
  context of suspended activities}.
\newblock \bibinfo{journal}{\emph{Human--Computer Interaction}}
  \bibinfo{volume}{32}, \bibinfo{number}{5-6} (\bibinfo{year}{2017}),
  \bibinfo{pages}{511--558}.
\newblock


\bibitem[Russell and Chi(2014)]%
        {russell2014looking}
\bibfield{author}{\bibinfo{person}{Daniel~M Russell} {and}
  \bibinfo{person}{Ed~H Chi}.} \bibinfo{year}{2014}\natexlab{}.
\newblock \showarticletitle{Looking back: Retrospective study methods for HCI}.
\newblock In \bibinfo{booktitle}{\emph{Ways of Knowing in HCI}}.
  \bibinfo{publisher}{Springer}, \bibinfo{pages}{373--393}.
\newblock


\bibitem[Russell and Oren(2009)]%
        {russell2009retrospective}
\bibfield{author}{\bibinfo{person}{Daniel~M Russell} {and}
  \bibinfo{person}{Mike Oren}.} \bibinfo{year}{2009}\natexlab{}.
\newblock \showarticletitle{Retrospective cued recall: a method for accurately
  recalling previous user behaviors}. In \bibinfo{booktitle}{\emph{2009 42nd
  Hawaii International Conference on System Sciences}}. IEEE,
  \bibinfo{pages}{1--9}.
\newblock


\bibitem[Safer and Murphy(2007)]%
        {safer2007comparing}
\bibfield{author}{\bibinfo{person}{Izzet Safer} {and} \bibinfo{person}{Gail~C
  Murphy}.} \bibinfo{year}{2007}\natexlab{}.
\newblock \showarticletitle{Comparing episodic and semantic interfaces for task
  boundary identification}. In \bibinfo{booktitle}{\emph{Proceedings of the
  2007 conference of the center for advanced studies on Collaborative
  research}}. \bibinfo{pages}{229--243}.
\newblock


\bibitem[Sellen et~al\mbox{.}(2007)]%
        {sellen2007life}
\bibfield{author}{\bibinfo{person}{Abigail~J Sellen}, \bibinfo{person}{Andrew
  Fogg}, \bibinfo{person}{Mike Aitken}, \bibinfo{person}{Steve Hodges},
  \bibinfo{person}{Carsten Rother}, {and} \bibinfo{person}{Ken Wood}.}
  \bibinfo{year}{2007}\natexlab{}.
\newblock \showarticletitle{Do life-logging technologies support memory for the
  past? An experimental study using SenseCam}. In
  \bibinfo{booktitle}{\emph{Proceedings of the SIGCHI conference on Human
  factors in computing systems}}. \bibinfo{pages}{81--90}.
\newblock


\bibitem[Shen et~al\mbox{.}(2008)]%
        {shen2008automatically}
\bibfield{author}{\bibinfo{person}{Jianqiang Shen}, \bibinfo{person}{Werner
  Geyer}, \bibinfo{person}{Michael Muller}, \bibinfo{person}{Casey Dugan},
  \bibinfo{person}{Beth Brownholtz}, {and} \bibinfo{person}{David~R Millen}.}
  \bibinfo{year}{2008}\natexlab{}.
\newblock \showarticletitle{Automatically finding and recommending resources to
  support knowledge workers' activities}. In
  \bibinfo{booktitle}{\emph{Proceedings of the 13th international conference on
  Intelligent user interfaces}}. \bibinfo{pages}{207--216}.
\newblock


\bibitem[Smith et~al\mbox{.}(2003)]%
        {smith2003groupbar}
\bibfield{author}{\bibinfo{person}{Greg Smith}, \bibinfo{person}{Patrick
  Baudisch}, \bibinfo{person}{George Robertson}, \bibinfo{person}{Mary
  Czerwinski}, \bibinfo{person}{Brian Meyers}, \bibinfo{person}{Daniel
  Robbins}, {and} \bibinfo{person}{Donna Andrews}.}
  \bibinfo{year}{2003}\natexlab{}.
\newblock \showarticletitle{Groupbar: The taskbar evolved}. In
  \bibinfo{booktitle}{\emph{Proceedings of OZCHI}}, Vol.~\bibinfo{volume}{3}.
  \bibinfo{pages}{10}.
\newblock


\bibitem[Standing et~al\mbox{.}(1970)]%
        {standing1970perception}
\bibfield{author}{\bibinfo{person}{Lionel Standing}, \bibinfo{person}{Jerry
  Conezio}, {and} \bibinfo{person}{Ralph~Norman Haber}.}
  \bibinfo{year}{1970}\natexlab{}.
\newblock \showarticletitle{Perception and memory for pictures: Single-trial
  learning of 2500 visual stimuli}.
\newblock \bibinfo{journal}{\emph{Psychonomic science}} \bibinfo{volume}{19},
  \bibinfo{number}{2} (\bibinfo{year}{1970}), \bibinfo{pages}{73--74}.
\newblock


\bibitem[Swearngin et~al\mbox{.}(2021)]%
        {swearngin2021scraps}
\bibfield{author}{\bibinfo{person}{Amanda Swearngin}, \bibinfo{person}{Shamsi
  Iqbal}, \bibinfo{person}{Victor Poznanski}, \bibinfo{person}{Mark
  Encarnaci{\'o}n}, \bibinfo{person}{Paul~N Bennett}, {and}
  \bibinfo{person}{Jaime Teevan}.} \bibinfo{year}{2021}\natexlab{}.
\newblock \showarticletitle{Scraps: Enabling Mobile Capture, Contextualization,
  and Use of Document Resources}. In \bibinfo{booktitle}{\emph{Proceedings of
  the 2021 CHI Conference on Human Factors in Computing Systems}}.
  \bibinfo{pages}{1--14}.
\newblock


\bibitem[Tashman and Edwards(2012)]%
        {10.1145/2147783.2147791}
\bibfield{author}{\bibinfo{person}{Craig Tashman} {and}
  \bibinfo{person}{W.~Keith Edwards}.} \bibinfo{year}{2012}\natexlab{}.
\newblock \showarticletitle{WindowScape: Lessons Learned from a Task-Centric
  Window Manager}.
\newblock \bibinfo{journal}{\emph{ACM Trans. Comput.-Hum. Interact.}}
  \bibinfo{volume}{19}, \bibinfo{number}{1}, Article \bibinfo{articleno}{8}
  (\bibinfo{date}{may} \bibinfo{year}{2012}), \bibinfo{numpages}{33}~pages.
\newblock
\showISSN{1073-0516}
\urldef\tempurl%
\url{https://doi.org/10.1145/2147783.2147791}
\showDOI{\tempurl}


\bibitem[Teevan et~al\mbox{.}(2014)]%
        {teevan2014selfsourcing}
\bibfield{author}{\bibinfo{person}{Jaime Teevan}, \bibinfo{person}{Daniel~J
  Liebling}, {and} \bibinfo{person}{Walter~S Lasecki}.}
  \bibinfo{year}{2014}\natexlab{}.
\newblock \showarticletitle{Selfsourcing personal tasks}.
\newblock In \bibinfo{booktitle}{\emph{CHI'14 Extended Abstracts on Human
  Factors in Computing Systems}}. \bibinfo{pages}{2527--2532}.
\newblock


\bibitem[van Solingen et~al\mbox{.}(1998)]%
        {van1998interrupts}
\bibfield{author}{\bibinfo{person}{Rini van Solingen}, \bibinfo{person}{Egon
  Berghout}, {and} \bibinfo{person}{Frank van Latum}.}
  \bibinfo{year}{1998}\natexlab{}.
\newblock \showarticletitle{Interrupts: just a minute never is}.
\newblock \bibinfo{journal}{\emph{IEEE software}} \bibinfo{volume}{15},
  \bibinfo{number}{5} (\bibinfo{year}{1998}), \bibinfo{pages}{97--103}.
\newblock


\bibitem[Voida et~al\mbox{.}(2008)]%
        {voida2008re}
\bibfield{author}{\bibinfo{person}{Stephen Voida}, \bibinfo{person}{Elizabeth~D
  Mynatt}, {and} \bibinfo{person}{W~Keith Edwards}.}
  \bibinfo{year}{2008}\natexlab{}.
\newblock \showarticletitle{Re-framing the desktop interface around the
  activities of knowledge work}. In \bibinfo{booktitle}{\emph{Proceedings of
  the 21st annual ACM symposium on User interface software and technology}}.
  \bibinfo{pages}{211--220}.
\newblock


\bibitem[Walton(2016)]%
        {doi:10.1080/13614533.2015.1137466}
\bibfield{author}{\bibinfo{person}{Graham Walton}.}
  \bibinfo{year}{2016}\natexlab{}.
\newblock \showarticletitle{“Digital Literacy” (DL): Establishing the
  Boundaries and Identifying the Partners}.
\newblock \bibinfo{journal}{\emph{New Review of Academic Librarianship}}
  \bibinfo{volume}{22}, \bibinfo{number}{1} (\bibinfo{year}{2016}),
  \bibinfo{pages}{1--4}.
\newblock
\urldef\tempurl%
\url{https://doi.org/10.1080/13614533.2015.1137466}
\showDOI{\tempurl}
\showeprint{https://doi.org/10.1080/13614533.2015.1137466}


\bibitem[Ware(2021)]%
        {WARE20211}
\bibfield{author}{\bibinfo{person}{Colin Ware}.}
  \bibinfo{year}{2021}\natexlab{}.
\newblock \showarticletitle{Chapter One - Foundations for an Applied Science of
  Data Visualization}.
\newblock In \bibinfo{booktitle}{\emph{Information Visualization (Fourth
  Edition)} (\bibinfo{edition}{fourth edition} ed.)},
  \bibfield{editor}{\bibinfo{person}{Colin Ware}} (Ed.).
  \bibinfo{publisher}{Morgan Kaufmann}, \bibinfo{pages}{1--29}.
\newblock
\showISBNx{978-0-12-812875-6}
\urldef\tempurl%
\url{https://doi.org/10.1016/B978-0-12-812875-6.00001-3}
\showDOI{\tempurl}


\bibitem[Weinreich et~al\mbox{.}(2006)]%
        {weinreich2006off}
\bibfield{author}{\bibinfo{person}{Harald Weinreich}, \bibinfo{person}{Hartmut
  Obendorf}, \bibinfo{person}{Eelco Herder}, {and} \bibinfo{person}{Matthias
  Mayer}.} \bibinfo{year}{2006}\natexlab{}.
\newblock \showarticletitle{Off the beaten tracks: exploring three aspects of
  web navigation}. In \bibinfo{booktitle}{\emph{Proceedings of the 15th
  international conference on World Wide Web}}. \bibinfo{pages}{133--142}.
\newblock


\bibitem[Williams et~al\mbox{.}(2019)]%
        {williams2019mercury}
\bibfield{author}{\bibinfo{person}{Alex~C Williams},
  \bibinfo{person}{Harmanpreet Kaur}, \bibinfo{person}{Shamsi Iqbal},
  \bibinfo{person}{Ryen~W White}, \bibinfo{person}{Jaime Teevan}, {and}
  \bibinfo{person}{Adam Fourney}.} \bibinfo{year}{2019}\natexlab{}.
\newblock \showarticletitle{Mercury: Empowering Programmers' Mobile Work
  Practices with Microproductivity}. In \bibinfo{booktitle}{\emph{Proceedings
  of the 32nd Annual ACM Symposium on User Interface Software and Technology}}.
  \bibinfo{pages}{81--94}.
\newblock


\end{thebibliography}










\appendix 
\section{Appendix}
\subsection{Semi-structured Interview Questions}

Q1: What was the most common reason when you create a screenshot in Scrapbook?\\
Q2: How did you use screenshots that you took in Scrapbook?\\
Q3: Based on your collection, could you please pick some screenshots and tell me why did you take them? Did Scrapbook help you recall past activities or retrieve digital resources?\\
Q4: What do you like about the Scrapbook and what do you not like about the Scrapbook? Advantages and disadvantages? \\
Q5: How do you curate your digital resources?  Compared to your own methods, how is Scrapbook?\\
Q6: Do you have any concerns about Scrapbook while using it?\\
Q7: Do you have any suggestions for Scrapbook or any features that you would like to see in our future iteration?

\end{document}